\documentclass[lettersize,journal]{IEEEtran}
\IEEEoverridecommandlockouts
\usepackage{cite}
\usepackage{amsmath,amssymb,amsfonts}
\usepackage{algorithmic}
\usepackage[linesnumbered,ruled,vlined]{algorithm2e}
\usepackage{graphicx}
\usepackage{textcomp}
\usepackage{xcolor}
\usepackage{caption}
\usepackage{subcaption}
\usepackage{mathrsfs}
\usepackage{epstopdf}
\usepackage{textcomp}
\usepackage{stfloats}
\usepackage{url}
\usepackage{verbatim}
\begin{document}

\title{Delay Alignment Modulation with Hybrid Analog/Digital Beamforming for Millimeter Wave and Terahertz Communications \\
\thanks{Part of this work has been presented at the IEEE WCNC 2023~\cite{DAM+HY}.}
}

\author{Jieni~Zhang,~Yong~Zeng,~\IEEEmembership{Senior~Member,~IEEE},~Xiangbin~Yu,~\IEEEmembership{Senior~Member,~IEEE},~Shi~Jin,~\IEEEmembership{Fellow,~IEEE}, Jinhong~Yuan,~\IEEEmembership{Fellow,~IEEE},~Ying-Chang~Liang,~\IEEEmembership{Fellow,~IEEE},~and~Rui~Zhang,~\IEEEmembership{Fellow,~IEEE}
\thanks{Jieni Zhang, Yong Zeng, and Shi Jin are with the National Mobile Communication Research Laboratory, Southeast University, Nanjing 210096, China; Yong Zeng is also with the Purple Mountain Laboratories, Nanjing 211111, China (e-mail: \{220230984, yong\_zeng, jinshi\}@seu.edu.cn). (\textit{Corresponding author: Yong Zeng.})}
\thanks{Xiangbin Yu is with the College of Electronic and Information Engineering, Nanjing University of Aeronautics and Astronautics, Nanjing 210016, China (e-mail: yxbxwy@gmail.com).}
\thanks{Jinhong Yuan is with the School of Electrical Engineering and Telecommunications, The University of New South Wales, Sydney, NSW 2052, Australia (e-mail: j.yuan@unsw.edu.au).}
\thanks{Ying-Chang Liang is with the National Key Laboratory of Wireless Communications, and the Center for Intelligent Networking and Communications (CINC), University of Electronic Science and Technology of China (UESTC), Chengdu 611731, China (e-mail: liangyc@ieee.org).}
\thanks{Rui Zhang is with the School of Science and Engineering, Shenzhen Research Institute of Big Data, The Chinese University of Hong Kong, Shenzhen, Guangdong 518172, China (e-mail: rzhang@cuhk.edu.cn).}
}
\maketitle
\begin{abstract}
For millimeter wave (mmWave) or Terahertz (THz) communications, by leveraging the high spatial resolution offered by large antenna arrays and the multi-path sparsity of mmWave/THz channels, a novel inter-symbol interference (ISI) mitigation technique called delay alignment modulation (DAM) has been recently proposed. The key ideas of DAM are \textit{delay pre-compensation} and \textit{path-based beamforming}. 
However, existing research on DAM is mainly based on fully digital beamforming, which requires the number of radio frequency (RF) chains to be equal to the number of antennas. 
This paper proposes the hybrid analog/digital beamforming based DAM, including both fully and partially connected structures. The analog and digital beamforming matrices are designed to achieve performance close to DAM based on fully digital beamforming. While DAM was considered for the path-based channel model with integer delays in the previous work, this paper extends DAM to a more general tap-based model that accounts for fractional path delays. To further reduce the cost of channel estimation and improve the performance for wireless channels with fractional delays, DAM with codebook-based beam alignment and DAM-orthogonal frequency division multiplexing (DAM-OFDM) with hybrid beamforming are proposed. The effectiveness of the proposed techniques is verified by extensive simulation results.

\end{abstract}


\section{Introduction}
Over the years, there has been an increasing demand for high-capacity data transmission, catalyzing extensive research into massive multiple-input multiple-output (MIMO) or even extremely large-scale MIMO (XL-MIMO) systems~\cite{XL-MIMO}, especially when integrated with millimeter-wave (mmWave) and Terahertz (THz) communication technologies~\cite{mmWave1,mmWave2}. These advanced communication paradigms are promising to significantly enhance data throughput and spectral efficiency. 
However, broadband communication systems face a critical challenge due to time-dispersive channels caused by multi-path propagation, leading to inter-symbol interference (ISI)~\cite{wireless1,wireless0}. Such interference degrades the overall system performance, causing signal distortion which deteriorates the reliability of data transmission. 

Numerous methods have been proposed to address the ISI issue, including time-domain and frequency-domain techniques.
Specifically, time-domain equalization techniques at the receiver are traditional methods for mitigating ISI, such as linear equalizers~\cite{MMSE,LE} and decision feedback equalization (DFE)~\cite{DFE2}.
Linear equalizers are designed to reverse the effects of the channel dispersion by applying the linear filter, but their complexity increases with longer channel delay spread.
DFE improves the equalization performance by incorporating previously detected symbols into the feedback loop, thereby reducing error rates. However, it is more complex to implement and its practical performance is sensitive to error propagation~\cite{DFE2}.
Precoding techniques, such as linear precoding and Tomlinson–Harashima (TH) precoding, can also be applied at the transmitter for eliminating ISI in the time domain~\cite{Precoding,THP}. Their basic ideas are similar to equalization at the receiver, whereas they are implemented at the transmitter side.
Besides, time reversal (TR) filters transmit signal with the time reverse of the channel impulse responses (CIR), so that multi-path signals are coherently combined at the receiver, thus effectively reducing ISI~\cite{TR1,TR2}. However, in MIMO systems, ISI can only be eliminated asymptotically as the number of the base station (BS) antennas goes to infinity, and the implementation of TR requires rate back-off techniques, which results in a reduction in spectral efficiency~\cite{TR3,TR4}.
RAKE receivers are commonly used in code-division multiple access (CDMA) systems, by applying multiple delayed matched filters to capture and combine signals from different paths coherently at the receiver~\cite{wireless1}. However, due to the use of spread spectrum technology, a much larger bandwidth than that of the transmitted information is required, and their spectral efficiency is limited. Moreover, integrating CDMA in MIMO systems incurs even higher complexity as additional spatial interference needs to be coped with~\cite{RAKE}.
An ISI suppression method using space time block coding (STBC) or space frequency block coding (SFBC) and statistical pre-filtering (SPF) was proposed in~\cite{STBC} and~\cite{SFBC}, where SPF combines transmit beamforming with signal pre-alignment. This approach not only achieves beamforming gain and ISI mitigation with SPF, but also achieves path diversity gain with STBC/SFBC simultaneously. However, the detailed designs provided in~\cite{STBC,SFBC,SFBC2} are applicable to wireless channels involving two paths only. Furthermore, in rich multi-path environments, where more than two beams are utilized and orthogonal STBC is applied to mitigate ISI, a rate penalty is incurred due to the coding rate being strictly less than one~\cite{STBC} for a large number of antennas.

On the other hand, frequency-domain equalization is an alternative way to address the ISI issue~\cite{FDE1,FDE2,FDE3}. It involves transforming the signal into the frequency domain, applying equalization, and then transforming it back to the time domain. This approach is particularly effective in combating ISI in wideband channels with frequency-domain modulation techniques and can be efficiently implemented using the fast Fourier transform (FFT). 
In particular, orthogonal frequency division multiplexing (OFDM) stands out among various frequency-domain ISI-mitigation modulation techniques for its robustness and efficiency~\cite{wireless1,OFDM}. By dividing the high-rate data stream into multiple parallel low-rate data streams over different sub-carriers, OFDM converts a frequency-selective channel into multiple frequency-flat channels, thus simplifying equalization while effectively eliminating ISI. However, OFDM also faces several critical issues in practical systems. For instance, the high peak-to-average-power ratio (PAPR) caused by the superposition of multiple sub-carrier signals in the time domain can lead to nonlinear distortion of power amplifiers~\cite{PAPR1,PAPR2}. Moreover, it suffers from severe out-of-band (OOB) emission that can cause adjacent channel interference~\cite{OOB}, and is sensitive to carrier frequency offset (CFO)~\cite{CFO}. More recently, new wideband modulation techniques such as orthogonal time-frequency-space (OTFS) modulation~\cite{OTFS,OTFS2,OTFS3} and orthogonal delay-Doppler division multiplexing (ODDM) modulation~\cite{ODDM} have attracted much attention, which perform modulation in the delay-Doppler domain. In addition to overcoming frequency-selective fading of the channel, they can counteract significant Doppler effects that occur in highly mobile environments. 
However, they also incur high implementation complexity and more signal processing latency.

In contrast to the aforementioned techniques, delay alignment modulation (DAM) was proposed recently as a new means to deal with ISI~\cite{DAM}. It builds upon the novel concepts of \textit{delay pre-compensation} and \textit{path-based beamforming}, which can achieve ISI mitigation and path diversity gain simultaneously. Note that unlike~\cite{STBC}, STBC is not required for DAM. In ideal cases, DAM can achieve optimal performance without the need for traditional equalization methods such as channel equalization or multi-carrier transmission.
Specifically, by exploiting the high spatial resolution offered by large antenna arrays \cite{spatial resolution} and leveraging the sparsity of multi-path propagation in mmWave/THz channels \cite{Sparsity,Sparsity-1,Sparsity-2}, DAM can completely remove the ISI by properly designing symbol delays at the transmitter to compensate for different delays of multi-path components in the channel. Subsequently, by employing path-based beamforming techniques, all multi-path signal components that reach the receiver simultaneously can be constructively combined.
As a result, this technique not only eliminates ISI at the receiver but also enables full utilization of the channel power contributed by all multi-path components.
Moreover, the authors in \cite{DAM-OFDM} proposed a more general DAM technique, aimed at reducing the channel delay spread when complete elimination of ISI is infeasible or undesirable. 
Furthermore, the concept of DAM-OFDM was proposed, offering the potential to reduce the cyclic prefix (CP) length and alleviate the high PAPR issue encountered in conventional OFDM.
The application of DAM in multi-user communication and wireless systems assisted by intelligent reflecting surfaces (IRSs) was studied in \cite{DAM-mu1,DAM-mu2,DAM-IRS}. 
In addition, an efficient channel estimation technique for DAM was proposed in \cite{DAM-CE}. 
Furthermore, \cite{fDAM} investigated DAM for the more general and practical scenarios with fractional multi-path delays. Integrated sensing and communications (ISAC) based on DAM were investigated in~\cite{DAM-ISAC1,DAM-ISAC2,DAM-ISAC3,DAM-ISAC4}. In addition, \cite{ISAC-IRS1} focused on a THz ISAC system that employs active reconfigurable intelligent surfaces (RIS) and DAM, while the secure transmission problem for such a system was later studied in~\cite{ISAX-IRS2}.
To further overcome the Doppler effects in the high-mobility communication, delay-Doppler alignment modulation (DDAM) was proposed by leveraging the \textit{delay-Doppler compensation} and \textit{path-based beamforming}~\cite{DDAM}, which is a generalization of DAM. Compared to OFDM and OTFS, DDAM offers lower PAPR, higher spectral efficiency, and lower complexity at both the transmitter and receiver, especially in scenarios with fewer multi-paths and a large number of information-bearing symbols transmitted per channel coherence block~\cite{DDAM2}. Additionally, it can be combined with other waveform design techniques to improve their performance, such as DDAM-OFDM and DDAM-OTFS~\cite{DDAM2,DDAM-OTFS}.

The aforementioned works have demonstrated the great potential of DAM for spatially sparse channels with large antenna arrays, for enhancing spectral efficiency, reducing PAPR, and simplifying receiver design. 
However, the previous works on DAM or its extensions have assumed fully digital beamforming, which may incur substantial hardware cost and power consumption since the required number of radio frequency (RF) chains needs to be equal to the number of antennas.
To address this issue, significant research endeavors have been devoted to the investigation of hybrid analog/digital beamforming, which greatly reduces the number of RF chains for connecting the digital beamformer and analog beamformer~\cite{Hybrid-ana,OMP,Alt}. 

Therefore, in this paper, we propose the DAM architecture based on hybrid analog/digital beamforming~\cite{DAM+HY}, which aims to achieve DAM performance comparable to that of fully digital beamforming, but with a smaller number of RF chains. 
This paper thus aims to explore how DAM can be realized based on hybrid analog/digital beamforming, thus providing new practical designs to further reduce the power consumption and hardware cost for implementing DAM. The main contributions of this paper are summarized as follows:
\begin{itemize}
	\item This paper introduces a new architecture of DAM using hybrid analog/digital beamforming, which includes both fully connected and partially connected structures. In our preliminary work~\cite{DAM+HY}, DAM was considered for the path-based channel model with integer delays. In contrast, this paper further explores DAM based on the more general tap-based channel model with fractional delays~\cite{DAM-CE,fDAM}.
	\item To reduce the overhead and complexity associated with channel estimation, a new method for implementing DAM via codebook-based beam alignment is proposed. Specifically, the analog beamforming vectors are determined first by searching the discrete Fourier transform (DFT) codebook, which helps in aligning the beams efficiently. After obtaining the analog beamforming matrix, to determine the digital beamforming matrix, only the equivalent channel needs to be estimated, which has a dimension much smaller than the original channel dimension.
	\item Under the tap-based channel model with fractional delays, where taps within the same cluster may exhibit strong correlation and thus the conventional ISI-zero-forcing (ISI-ZF) beamforming~\cite{DAM} becomes ineffective, DAM-OFDM based on hybrid beamforming is thus proposed. This method aligns each cluster using DAM and then uses OFDM to further address the residual ISI that has much smaller channel spread. This not only leverages the advantage of DAM in reducing delay spread to significantly reduce the number of OFDM sub-carriers or CP overhead, but also takes advantage of OFDM's capability for flexible time-frequency resource allocations.
	\item The effectiveness of the proposed methods, including hybrid beamforming based DAM, beam alignment based DAM, and hybrid beamforming based DAM-OFDM, is validated through simulations. 
\end{itemize}

The rest of this paper is organized as follows. Section~\ref{System Model} presents the system model, where both the path-based and tap-based channel models are introduced. The key ideas of DAM based on hybrid beamforming with fully connected and partially connected structures are detailed in Section~\ref{HB-DAM}. Section~\ref{BA-DAM} proposes the beam alignment based DAM to reduce the channel estimation overhead. DAM-OFDM based on hybrid beamforming to overcome the challenges in channels with fractional delays is presented in Section~\ref{HB-DAM-OFDM}. Section~\ref{Simulation} provides the simulation results and Section~\ref{Conclusion} concludes the paper.

\emph{Notations:}
Scalars are denoted by italic letters, and for real numbers $x$ and $y$, $\lceil x \rceil$ denotes ceiling operations on $x$, and $\text{mod}(x,y)$ returns the remainder on dividing $x$ by $y$. 
Vectors are denoted by boldface lower-case letters, and for a vector $\mathbf{x}$, $\|\mathbf{x}\|$ denotes its $l_2$-norm.
Matrices are denoted by boldface upper-case letters. $\mathbf{X}^T$, $\mathbf{X}^{*}$, $\mathbf{X}^H$, $\mathbf{X}^\dagger$ and $\|\mathbf{X}\|_F$ denote the transpose, conjugate, Hermitian transpose, pseudo-inverse and Frobenius norm of matrix $\mathbf{X}$, respectively.
The block diagonal matrix is denoted by $\text{diag}(\mathbf{x}_1,\mathbf{x}_2,\cdots,\mathbf{x}_n)$, where each block is a column vector and its diagonal elements are $\mathbf{x}_i, i=1,\cdots,n$.
Sets are denoted by capital calligraphy letters, and for a set $\mathcal{X}$, $|\mathcal{X}|$ denotes its cardinality.
$\mathbb{C}^{M \times N}$ denotes the space of $M \times N$ complex-valued matrices.
The imaginary unit of complex numbers is denoted by the symbol $j$, with $j^2=-1$.
$\mathbb{E} (\cdot)$ and $*$ denote the statistical expectation and the linear convolution operation, respectively.
${\cal C}{\cal N}(\mu,{\sigma ^2})$ denotes the distribution of a circularly symmetric complex Gaussian (CSCG) random variable with expected value $\mu$ and variance $\sigma^2$, and $\sim $ stands for ``distributed as''.
$\delta[\cdot]$ denotes the Kronecker delta function.

\section{System Model} \label{System Model}

\begin{figure}[t!]
	\centerline{\includegraphics[width=7cm]{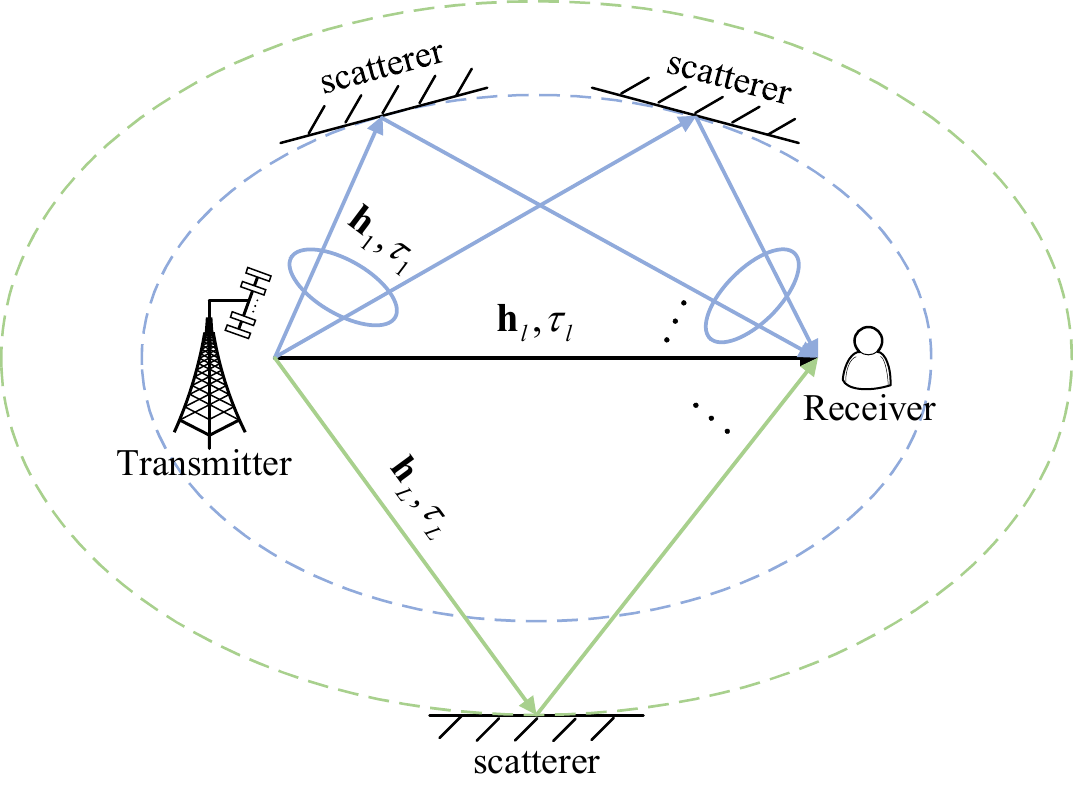}}
	\caption{A MISO spatially sparse mmWave/THz communication system with fractional multi-path delays~\cite{fDAM}.}
	\label{MISO}
\end{figure}

As shown in Fig.~\ref{MISO}, we consider a spatially sparse wireless communication system, such as mmWave or THz system, in which a base station (BS) equipped with  ${M_\mathrm{t}} \gg 1$ antennas engages in communication with a single-antenna user equipment (UE). 
For cost-effective implementation, the BS has only ${M_{\mathrm{RF}}} < {M_\mathrm{t}}$ RF chains, and hybrid analog/digital beamforming is applied.
We assume a quasi-static block fading environment, where the channel remains constant within each coherent block and may vary across blocks.

Considering the path-based channel model used in the previous work~\cite{DAM+HY}, the discrete-time representation of the downlink channel impulse response is
\begin{equation}
	{{\mathbf{h}}_{\mathrm{DL}}^H}[n] = \sum\nolimits_{l = 1}^L {{\mathbf{h}}_l^H} \delta [n - {n_l}],
	\label{Channel1}
\end{equation}
where $L$ represents the number of temporally resolvable multi-paths, ${{\mathbf{h}}_l} \in \mathbb{C}^{{M_\mathrm{t}} \times 1}$ and $n_l$ represent the channel vector and the discretized delay of the $l$th multi-path, respectively. Let the minimum delay among all $L$ multi-paths be denoted by ${n_{\min }} \buildrel \Delta \over = \mathop {\min }\limits_{1 \le l \le L} {n_l}$, and the maximum delay be denoted by ${n_{\max }} \buildrel \Delta \over = \mathop {\max }\limits_{1 \le l \le L} {n_l}$. Thus, ${n_{{\rm{span}}}} = {n_{\max }} - {n_{\min }}$ represents the normalized channel delay spread. Note that multi-path delays in the channel model of \eqref{Channel1} are assumed to be integer multiples of sampling interval $T_{\mathrm{s}}$.
However, the delay of multi-paths may be non-integer multiples of $T_{\mathrm{s}}$ in reality. To generalize our approach and facilitate channel estimation in~\cite{DAM-CE}, the tap-based channel model should be considered further.
Considering the tap-based channel model, the $q$th delay tap of the downlink baseband channel can be expressed as~\cite{DAM-CE}
\begin{equation}
	{\textbf{h}}_{\mathrm{DL}}^H[q] = \sum\nolimits_{l = 1}^L {{\textbf{h}}_l^Hp(q{T_{\mathrm{s}}} - {\tau_l}),q = 0, \ldots ,Q},
	\label{channel}
\end{equation}
where $p(t)$ is the pulse shaping function and $\tau_l$ represents the physical  delay of the $l$th multi-path. Note that the multi-path delays $\tau_l$ do not have to be integer multiples of $T_{\mathrm{s}}$. The total number of delay taps is $Q+1=\left\lceil\tau_{\mathrm{UB}} /T_{\mathrm{s}}\right\rceil+1$, where $\tau_{\mathrm{UB}}$ is a sufficiently large delay value beyond which no significant power can be received.
Due to the sparsity of multi-paths in mmWave/THz communications using large antenna arrays, we usually have $L \ll (Q+1)$ and $L \ll {M_\mathrm{t}}$~\cite{Sparsity-1,Sparsity-2,Sparsity}. In this case, out of the $(Q+1)$ channel taps, only around $L$ taps have significant power.
Because each temporally resolvable multi-path in \eqref{Channel1} and \eqref{channel} may comprise multiple sub-paths with the same delay but distinct angles of departure (AoDs), ${\mathbf{h}}_l$ can be further expressed as~\cite{DAM}
\begin{equation}
	{{\bf{h}}_l} = {\alpha _l}\sum\nolimits_{i = 1}^{{\mu _l}} {{v_{li}}{{\bf{a}}_\mathrm{t}}\left( {{\theta _{li}}} \right)}, 
	\label{hl}
\end{equation}
where $\alpha_l$ represents the complex-valued path gain of the $l$th multi-path, $\mu _l$ represents the number of sub-paths within the $l$th multi-path, ${v_{li}} = \sqrt {{\varsigma _{li}}} {e^{j{\phi _{li}}}}$ represents the complex coefficient of the $i$th sub-path of the $l$th multi-path, where $\phi _{li}$ denotes the phase of the $i$th sub-path, and $\varsigma _{li}$ represents the power allocation for the $i$th sub-path and satisfies the condition $\sum\nolimits_{i = 1}^{{\mu _l}} {{\varsigma _{li}} = 1} $ to ensure the power is normalized, and $\theta_{li}$ denotes the AoD of the $i$th sub-path of the $l$th multi-path, while ${{\mathbf{a}}_\mathrm{t}}({\theta _{li}}){ \in \mathbb{C}^{{M_\mathrm{t}} \times 1}}$ denotes  the corresponding transmit array response vector. For a basic uniform linear array (ULA), ${{\mathbf{a}}_\mathrm{t}}({\theta _{li}})$ is given by
\begin{equation}
	{{\mathbf{a}}_\mathrm{t}}({\theta _{li}}) = {\left[ {1,{e^{ - j\frac{{2\pi d}}{\lambda }\sin ({\theta _{li}})}}, \cdots ,{e^{ - j\frac{{2\pi ({M_\mathrm{t}} - 1)d}}{\lambda }\sin ({\theta _{li}})}}} \right]^T},
	\label{array response}
\end{equation}
where $\lambda$  denotes the signal wavelength and $d$ denotes the inter-element spacing of the ULA.

Let ${\mathbf{x}}[n]{ \in \mathbb{C}^{{M_\mathrm{t}} \times 1}}$ denote the discrete-time representation of the signals transmitted by the BS. 
Based on the path-based channel model in \eqref{Channel1}, the received signal at UE is
\begin{equation}
	y[n] = {{\bf{h}}_{\mathrm{DL}}^H}[n] * {\bf{x}}[n] + z[n] = \sum\limits_{l = 1}^L {{\bf{h}}_l^H} {\bf{x}}\left[ {n - {n_l}} \right] + z[n],
	\label{y[n]p}
\end{equation}
where $z[n] \sim {\cal C}{\cal N}(0,{\sigma ^2})$ represents the additive white Gaussion noise (AWGN). Similarly, based on the tap-based channel model, the received signal can be expressed as 
\begin{equation}
	y[n]=\sum\nolimits_{q=0}^Q \mathbf{h}_{\mathrm{DL}}^H[q] \mathbf{x}[n-q]+z[n].
	\label{y[n]t}
\end{equation}
It can be seen from both \eqref{y[n]p} and \eqref{y[n]t} that in a multipath propagation environment, the received signal is a superposition of the transmitted signal with different delays, which can cause ISI. In~\cite{DAM}, a novel technique known as DAM was introduced to address the issue of ISI based on \textit{delay pre-compensation} and \textit{path-based beamforming}, obviating the need for conventional methods such as channel equalization or multi-carrier transmission. 
Yet, the DAM technique presented there assumes the fully digital beamforming at the BS, which cannot be applied for the considered system with fewer RF chains than the number of antennas. In the following, we propose the hybrid analog/digital beamforming based DAM technique for spatially sparse communication systems.

\section{DAM with Hybrid Beamforming} \label{HB-DAM}


\begin{figure}[!t]
	\centering
	\subfloat[Fully connected structure\label{DAM1}]{
		\includegraphics[width=8.5cm]{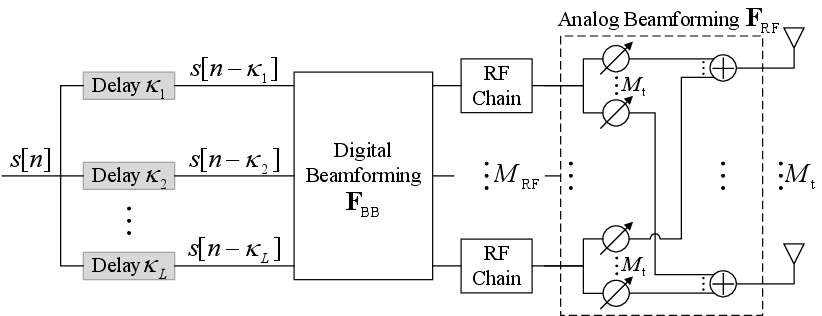}}\\
	\subfloat[Partially connected structure\label{DAM_p}]{
		\includegraphics[width=8.5cm]{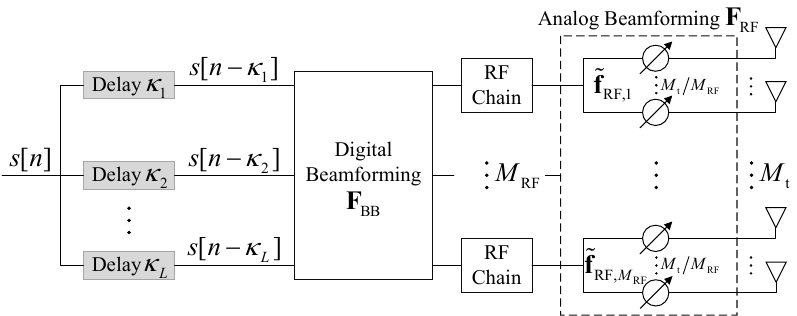}}	
	\caption{Transmitter architecture for DAM based on hybrid beamforming with fully connected structure and partially connected structure.}
	\label{DAM1+p}
\end{figure}

The transmitter architecture of the proposed DAM system, utilizing hybrid analog/digital beamforming with the fully connected structure, is illustrated in Fig.~\ref{DAM1}.
This section will consider the implementation of DAM with hybrid beamforming based on path-based and tap-based channel models.
\subsection{Path-based DAM with Hybrid Beamforming}
Through \textit{delay pre-compensation} and \textit{path-based beamforming}, the transmitted signal for DAM with hybrid beamforming can be expressed as
\begin{equation}
	{\bf{x}}[n] = {{\bf{F}}_{\mathrm{RF}}}\sum\nolimits_{l = 1}^L {{{\bf{f}}_{\mathrm{BB},l}}s[n - {\kappa _l}]},
	\label{x[n]p}
\end{equation}
where ${{\bf{F}}_{\mathrm{RF}}}{ \in \mathbb{C}^{{M_\mathrm{t}} \times {M_{\mathrm{RF}}}}}$ denotes the analog beamforming matrix with unit modulus for each of its element, i.e., $| {{{\left( {{{\bf{F}}_{\mathrm{RF}}}} \right)}_{i,j}}} | = 1,\forall i,j$, ${{\bf{F}}_{\mathrm{BB}}}{ \in \mathbb{C}^{{M_{\mathrm{RF}}} \times L}}$ denotes the digital baseband beamforming matrix that includes ${{\bf{f}}_{\mathrm{BB},l}}{ \in \mathbb{C}^{{M_{\mathrm{RF}}} \times 1}}$ for its $l$th column, $s[n]$ is the independent and identically distributed (i.i.d.) information-bearing symbols satisfying the normalized power requirement $\mathbb{E}[ {{{| {s[ n ]} |}^2}} ] = 1$, and ${\kappa _l} \ge 0$ denotes the intentionally introduced delay with ${\kappa _l} \ne {\kappa _{l'}},\forall l \ne l'$, aiming to compensate for the delay $n_l$ of the $l$th channel path.

The power of the transmitted signal $\bf{x}\rm[n]$ is
\begin{equation}
	\begin{split}
		\mathbb{E}\left[ {\left\| {{\bf{x}}[n]} \right\|^2} \right] =& \sum\limits_{l = 1}^L {\mathbb{E}\left[ {\left\| {{{\bf{F}}_{\mathrm{RF}}}{{\bf{f}}_{\mathrm{BB},l}}s[n - {\kappa _l}]} \right\|^2} \right]} \\
		=& \sum\limits_{l = 1}^L {\left\| {{{\bf{F}}_{\mathrm{RF}}}{{\bf{f}}_{\mathrm{BB},l}}} \right\|^2 = \left\| {{{\bf{F}}_{\mathrm{RF}}}{{\bf{F}}_{\mathrm{BB}}}} \right\|_F^2 \le P} ,
	\end{split}
	\label{P}
\end{equation}
where $P$ represents the available transmit power. 
By substituting \eqref{x[n]p} into \eqref{y[n]p}, the received signal of DAM with hybrid analog/digital beamforming is
\begin{equation}
	\begin{split}
		y[n] = & \sum\limits_{l = 1}^L {{\bf{h}}_l^H{{\bf{F}}_{\mathrm{RF}}}{{\bf{f}}_{\mathrm{BB,}l}}s[n - {\kappa _l} - {n_l}] + } \\
		&\sum\limits_{l = 1}^L {\sum\limits_{l' \ne l}^L {{\bf{h}}_l^H{{\bf{F}}_{\mathrm{RF}}}{{\bf{f}}_{\mathrm{BB,}l'}}s[n - {\kappa _{l'}} - {n_l}]} }  + z[n].
	\end{split}
	\label{y[n] of DAM}
\end{equation}
According to the principle of DAM~\cite{DAM}, the introduced delay ${\kappa _l}$ is set to $	{\kappa _l} = {n_{\max }} - {n_l} \ge 0,\forall l$.
Thus, we have
\begin{equation}
	\begin{aligned}
		y[n] =&\left( {\sum\limits_{l = 1}^L {{\bf{h}}_l^H{{\bf{F}}_{\mathrm{RF}}}{{\bf{f}}_{\mathrm{BB,}l}}} } \right)s[n - {n_{\max }}] + \\
		&\sum\limits_{l = 1}^L {\sum\limits_{l' \ne l}^L {{\bf{h}}_l^H{{\bf{F}}_{\mathrm{RF}}}{{\bf{f}}_{\mathrm{BB,}l'}}s[n - {n_{\max }} + {n_{l'}} - {n_l}]} }  + z[n].
	\end{aligned}
	\label{y[n] with delay compensation}
\end{equation}
Observing \eqref{y[n] with delay compensation}, it becomes apparent that the first term contributes to the desired signal if the receiver is locked to the delay ${n_{\max }}$, whereas the second term represents the ISI.
Furthermore, the ISI can be eliminated by jointly designing $\bf{F}_{\mathrm{RF}}$ and $\bf{F}_{\mathrm{BB}}$ to satisfy
\begin{equation}
	{\bf{h}}_l^H{{\bf{F}}_{\mathrm{RF}}}{{\bf{f}}_{\mathrm{BB,}l'}} = 0,\forall l' \ne l.
	\label{ISI-ZF}
\end{equation}
Subsequently, the received signal in \eqref{y[n] with delay compensation} simplifies to
\begin{equation}
	y[n] = \left( {\sum\limits_{l = 1}^L {{\bf{h}}_l^H{{\bf{F}}_{\mathrm{RF}}}{{\bf{f}}_{\mathrm{BB,}l}}} } \right)s[n - {n_{\max }}] + z[n].
	\label{y[n] without ISI}
\end{equation}
It can be seen from \eqref{y[n] without ISI} that if condition in~\eqref{ISI-ZF} is satisfied, the ISI can be eliminated perfectly and the received signal becomes a symbol sequence with a single delay $n_{\max}$ and multiplied by the gains with $L$ multi-path contributions. The method in \eqref{ISI-ZF} is known as ISI-ZF beamforming. In this case, the SNR is
\begin{equation}
	\gamma  = \frac{1}{{{\sigma ^2}}}{\left| {\sum\nolimits_{l = 1}^L {{\bf{h}}_l^H{{\bf{F}}_{\mathrm{RF}}}{{\bf{f}}_{\mathrm{BB,}l}}} } \right|^2}.
	\label{SNR}
\end{equation}

As a result, the analog and digital beamforming matrices ${\bf{F}}_{\mathrm{RF}}$ and ${\bf{F}}_{\mathrm{BB}}$ need to be jointly designed to maximize the spectral efficiency of path-based DAM with the ISI-ZF constraint in \eqref{ISI-ZF}. The problem can be stated as
\begin{equation}
	\begin{split}
		\mathop {\max }\limits_{{{\bf{F}}_{\mathrm{RF}}},{{\bf{F}}_{\mathrm{BB}}}} \enspace&\log_2\left({1+\frac{1}{{{\sigma ^2}}}\left| {\sum\nolimits_{l = 1}^L {{\bf{h}}_l^H{{\bf{F}}_{\mathrm{RF}}}{{\bf{f}}_{\mathrm{BB,}l}}} } \right|^2}\right)\\
		\text{s.t.}\enspace&{\bf{h}}_l^H{{\bf{F}}_{\mathrm{RF}}}{{\bf{f}}_{\mathrm{BB,}l'}} = 0,\forall l' \ne l,\\
		&\left| {{{\left( {{{\bf{F}}_{\mathrm{RF}}}} \right)}_{i,j}}} \right| = 1,\forall i,j,
		\left\| {{{\bf{F}}_{\mathrm{RF}}}{{\bf{F}}_{\mathrm{BB}}}} \right\|_F^2 \le P.
	\end{split}
	\label{DAM with Hybrid_0}
\end{equation}
Apparently, the spectral efficiency can be maximized by maximizing the SNR in~\eqref{SNR}. By discarding the constant term, the problem in~\eqref{DAM with Hybrid_0} can be equivalently stated as
\begin{equation}
	\begin{split}
		\mathop {\max }\limits_{{{\bf{F}}_{\mathrm{RF}}},{{\bf{F}}_{\mathrm{BB}}}} \enspace&{\left| {\sum\nolimits_{l = 1}^L {{\bf{h}}_l^H{{\bf{F}}_{\mathrm{RF}}}{{\bf{f}}_{\mathrm{BB,}l}}} } \right|^2}\\
		\text{s.t.}\enspace&{\bf{h}}_l^H{{\bf{F}}_{\mathrm{RF}}}{{\bf{f}}_{\mathrm{BB,}l'}} = 0,\forall l' \ne l,\\
		&\left| {{{\left( {{{\bf{F}}_{\mathrm{RF}}}} \right)}_{i,j}}} \right| = 1,\forall i,j,
		\left\| {{{\bf{F}}_{\mathrm{RF}}}{{\bf{F}}_{\mathrm{BB}}}} \right\|_F^2 \le P.
	\end{split}
	\label{DAM with Hybrid}
\end{equation}
Directly solving the optimization problem \eqref{DAM with Hybrid} is difficult, since it is non-convex and the analog and digital beamforming vectors are closely coupled. By following the similar idea of hybrid beamforming optimization as \cite{OMP}, we may first determine the optimal solution for fully digital beamforming design, and then find the hybrid analog/digital beamforming matrices to closely approximate the optimal fully digital beamforming. Specifically, by letting  ${{\bf{F}}_{\mathrm{RF}}}{{\bf{f}}_{\mathrm{BB,}l}} = {{\bf{f}}_l},\forall l$, a new beamforming problem for DAM can be formulated as
\begin{equation}
	\begin{split}
		\mathop {\max }\limits_{\left\{ {{\bf{f}}_l} \right\}_{l = 1}^L} \enspace&{\left| {\sum\nolimits_{l = 1}^L {{\bf{h}}_l^H{{\bf{f}}_l}} } \right|^2}\\
		\text{s.t.}\enspace&{\bf{h}}_l^H{{\bf{f}}_{l'}} = 0,\forall l' \ne l,
		\sum\limits_{l = 1}^L {{{\left\| {{{\bf{f}}_l}} \right\|}^2}}  \le P.
	\end{split}
	\label{DAM with digital}
\end{equation}

Note that problem \eqref{DAM with digital} corresponds to ISI-ZF for DAM with fully digital beamforming, whose optimal solution $\{{\bf{f}}_l^{\mathrm{opt}}\}_{l=1}^L$ has been obtained in closed-form in \cite{DAM}. 

In order to achieve the same performance as fully digital beamforming based DAM, ${{\bf{F}}_{\mathrm{RF}}}$ and ${{\bf{F}}_{\mathrm{BB}}}$ should be jointly designed to satisfy
\begin{equation}
	{{\bf{F}}_{\mathrm{RF}}}{{\bf{f}}_{\mathrm{BB,}l}} = {\bf{f}}_l^{\mathrm{opt}},\forall l.
	\label{hybrid design condition}
\end{equation}

By letting ${{\bf{F}}_{\mathrm{opt}}} = [{\bf{f}}_1^{\mathrm{opt}},{\bf{f}}_2^{\mathrm{opt}}, \ldots {\bf{f}}_L^{\mathrm{opt}}]{ \in \mathbb{C}^{{M_\mathrm{t}} \times L}}$, \eqref{hybrid design condition} can be compactly expressed as: ${{\bf{F}}_{\mathrm{RF}}}{{\bf{F}}_{\mathrm{BB}}}{\rm{ = }}{{\bf{F}}_{\mathrm{opt}}}$. Note that the rank of ${\bf{F}}_{\mathrm{opt}}$ is $L$ when the vectors ${\bf{h}}_l,\forall l$ are linearly independent~\cite{DAM+HY}. On the other hand, the rank of ${{\bf{F}}_{\mathrm{RF}}}{{\bf{F}}_{\mathrm{BB}}}$ is no greater than ${M_{\mathrm{RF}}}$. This indicates that for \eqref{hybrid design condition} to hold, one necessary condition is ${M_{\mathrm{RF}}} \ge L$~\cite{M_RF,M_RF2}. 
Furthermore, it was revealed in \cite{M_RF2} that when ${M_{\mathrm{RF}}} \geq 2L$, ${{\bf{F}}_{\mathrm{RF}}}$ and ${{\bf{F}}_{\mathrm{BB}}}$ can be found to satisfy ${{\bf{F}}_{\mathrm{RF}}}{{\bf{F}}_{\mathrm{BB}}}{\rm{ = }}{{\bf{F}}_{\mathrm{opt}}}$ exactly, indicating that hybrid beamforming based DAM can fully realize the performance of fully digital beamforming when ${M_{\mathrm{RF}}} \ge 2L$.

With the optimal ISI-ZF beamforming matrix $\mathbf{F}_\mathrm{opt}$ for fully digital beamforming  obtained, hybrid beamforming for DAM can be designed by considering the optimization problem as
\begin{equation}
	\begin{split}
		\mathop {\min}\limits_{{{\bf{F}}_{\mathrm{RF}}}{\rm{,}}{{\bf{F}}_{\mathrm{BB}}}} \enspace &{\left\| {{{\bf{F}}_{\mathrm{opt}}} - {{\bf{F}}_{\mathrm{RF}}}{{\bf{F}}_{\mathrm{BB}}}} \right\|_F}\\
		\text{s.t.} \enspace &\left| {{{\left( {{{\bf{F}}_{\mathrm{RF}}}} \right)}_{i,j}}} \right| = 1,\forall i,j,
		\left\| {{{\bf{F}}_{\mathrm{RF}}}{{\bf{F}}_{\mathrm{BB}}}} \right\|_F^2 = P.
	\end{split}
	\label{hybrid problem 1}
\end{equation}
Our previous work in~\cite{DAM+HY} has proposed a solution for this problem, which transforms it into the following optimization problem
\begin{equation}
	\begin{split}
		\mathop {\min}\limits_{{{\bf{F}}_{\mathrm{RF}}}{\rm{,}}{{\bf{F}}_{\mathrm{BB}}}} \enspace&\left\| {{{\bf{F}}_{\mathrm{opt}}} - {{\bf{F}}_{\mathrm{RF}}}{{\bf{F}}_{\mathrm{BB}}}} \right\|_F\\
		\text{s.t.}\enspace&{{\bf{f}}_{\mathrm{RF},j}} \in {{\bf{a}}_\mathrm{t}}({\theta _{li}}),\forall l,i,j,
		{\left\| {{{\bf{F}}_{\mathrm{RF}}}{{\bf{F}}_{\mathrm{BB}}}} \right\|_F^2} = P.
	\end{split}
	\label{hybrid problem 2}
\end{equation}
This problem can be solved with the orthogonal matching pursuit (OMP) algorithm in~\cite{OMP}, which is omitted here for brevity.

Note that the above optimization problem considers a hybrid analog/digital beamforming design based on the fully connected structure as shown in Fig.~\ref{DAM1}. In the fully connected hybrid beamforming structure, each RF chain is connected to all the antenna elements. This structure is highly flexible, capable of achieving finer beam control and higher array gain, but it brings higher complexity and power consumption, since the total number of analog phase shifters required is $M_\mathrm{RF}M_\mathrm{t}$. For further cost and power savings, hybrid beamforming based on the partially connected structure can be considered, as shown in Fig.~\ref{DAM_p}. In the partially connected structure, each RF chain is connected to only a subset of the antenna elements, adding another constraint to the analog beamforming matrix ${{\bf{\tilde F}}_{\mathrm{RF}}}$, i.e., only a portion of the elements of ${{\bf{\tilde F}}_{\mathrm{RF}}}$ can be designed, with the remaining elements being zero. Specifically, assuming that $M={M_{\mathrm{t}}}/{M_{\mathrm{RF}}}$ antennas are connected to each RF chain, the analog beamforming matrix can be expressed as ${{\bf{\tilde F}}_{\mathrm{RF}}} = \mathrm{diag}({{\bf{\tilde f}}}_{\mathrm{RF,1}}, {{\bf{\tilde f}}}_{\mathrm{RF,2}}, \ldots, {{\bf{\tilde f}}}_{\mathrm{RF,}{M_{\mathrm{RF}}}})$, where ${{\bf{\tilde{f}}}_{\mathrm{RF},t}}{ \in \mathbb{C}^{M\times 1}}$ denotes the beamforming vector for the $t$th RF chain and $M$ is assumed to be an integer for notational convenience. As a result, the optimization problem for hybrid beamforming based DAM with the partially connected structure is given by
\begin{equation}
	\begin{aligned}
		& \min_{\mathbf{\tilde F}_{\mathrm{RF}}, \mathbf{F}_{\mathrm{BB}}} && \left\| {{{\bf{F}}_{\mathrm{opt}}} - {{\bf{\tilde F}}_{\mathrm{RF}}}{{\bf{F}}_{\mathrm{BB}}}} \right\|_F \\
		& \quad \quad \text{s.t.} && \mathbf{\tilde{f}}_{\mathrm{RF}, t} \in \mathbf{a}_\mathrm{t} \left( \theta_{li} \right)_{1+(t-1)M : tM}, \ \forall t, l, i, \\
		& && \left\| \mathbf{\tilde F}_{\mathrm{RF}} \mathbf{F}_{\mathrm{BB}} \right\|_F^2 = P.
	\end{aligned}
	\label{hybrid problem p0}
\end{equation}
According to the method in~\cite{Alt}, the power constraint in~\eqref{hybrid problem p0} can be temporally removed. So this problem can be decoupled into multiple single-objective optimization problems as
\begin{equation}
	\begin{aligned}
		& \min_{\mathbf{\tilde{f}}_{\mathrm{RF}, t}, \left( \mathbf{F}_{\mathrm{BB}} \right)_{t,:}} && \left\| \left( \mathbf{F}_{\text{opt}} \right)_{1+(t-1)M : tM, :} - \mathbf{\tilde{f}}_{\mathrm{RF}, t} \left( \mathbf{F}_{\mathrm{BB}} \right)_{t,:} \right\|_F, \ \forall t, \\
		& \quad \quad \text{s.t.} && \mathbf{\tilde{f}}_{\mathrm{RF}, t} \in \mathbf{a}_t \left( \theta_{li} \right)_{1+(t-1)M : tM}, \ \forall t, l, i.
	\end{aligned}
	\label{hybrid problem p}
\end{equation}
These problem can also be solved using the OMP algorithm in~\cite{OMP}, and the solutions are denoted by $\mathbf{\tilde F}^{\mathrm{opt}}_{\mathrm{RF}}$ and $\mathbf{F}^{\mathrm{opt}}_{\mathrm{BB}}$. Finally, to satisfy the power constraint in~\eqref{hybrid problem p0}, $\mathbf{F}^{\mathrm{opt}}_{\mathrm{BB}}$ needs to be normalized by a factor of $\sqrt{P}/\| \mathbf{\tilde F}^{\mathrm{opt}}_{\mathrm{RF}} \mathbf{F}^{\mathrm{opt}}_{\mathrm{BB}} \|_F$. 

Because of the unit modulus constraint imposed on the analog beamforming matrix, the performance of hybrid beamforming based on the fully connected structure may be different from the optimal fully digital beamforming when ${M_{\mathrm{RF}}} < 2L$, and the performance based on partially connected structures is even worse. In other words, with the aforementioned hybrid analog/digital beamforming design, there is no guarantee that the ISI-ZF constraint in~\eqref{ISI-ZF} can be satisfied. This implies that for the purpose of evaluating the resulting performance of DAM based on hybrid analog/digital beamforming, the residual ISI needs to be considered.

To this end, the same delay components in \eqref{y[n] with delay compensation} should be grouped together~\cite{DAM}. Specifically, let ${\cal L} \buildrel \Delta \over = \left\{ {l:l = 1, \ldots ,L} \right\}$ denote the set of all multi-paths, and ${{\cal L}_l} \buildrel \Delta \over = {\cal L}\backslash l$ represent the subset of set ${\cal L}$ by removing the $l$th path from it.
Furthermore, let the delay difference between $l'$ and $l$ be denoted by ${\Delta _{l',l}} \buildrel \Delta \over = {n_{l'}} - {n_l}$. Then, for $\forall l \ne l'$, ${\Delta _{l',l}} \in \left\{ { \pm 1, \ldots , \pm {n_{\mathrm{span}}}} \right\}$. Thus, \eqref{y[n] with delay compensation} can be equivalently expressed as
\begin{equation}
	\begin{split}
		\hspace{-1ex} y[n] = &\left( {\sum\limits_{l = 1}^L {{\bf{h}}_l^H{{\bf{F}}_{\mathrm{RF}}}{{\bf{f}}_{\mathrm{BB,}l}}} } \right)s[n - {n_{\max }}] + \\
		&\sum\limits_{l = 1}^L {\sum\limits_{l' \ne l}^L {{\bf{h}}_l^H{{\bf{F}}_{\mathrm{RF}}}{{\bf{f}}_{\mathrm{BB,}l'}}s[n - {n_{\max }} + {\Delta _{l',l}}]} }  + z[n].
	\end{split}
	\label{y[n] with delta}
\end{equation}
The terms with the same delay difference in \eqref{y[n] with delta} correspond to the same symbols, which need to be grouped. For each delay difference $i \in \left\{ { \pm 1, \ldots , \pm {n_{\mathrm{span}}}} \right\}$, the effective channel can be defined as
\begin{equation}
	{\bf{g}}_{l'}^H[i] \buildrel \Delta \over = \left\{ \begin{array}{l}
		{\bf{h}}_l^H, \text{if} \enspace \exists l \in {{\cal L}_{l'}},\enspace \text{s.t.}\enspace{n_{l'}} - {n_l} = i,\\
		{\bf{0}},\text{otherwise}.
	\end{array} \right.
\end{equation}
Therefore, \eqref{y[n] with delta} can be expressed equivalently as
\begin{equation}
	\begin{aligned}
		y[n] = &\left( {\sum\limits_{l = 1}^L {{\bf{h}}_l^H{{\bf{F}}_{\mathrm{RF}}}{{\bf{f}}_{\mathrm{BB,}l}}} } \right)s[n - {n_{\max }}] + \\
		&\sum\limits_{i =  - {n_{\mathrm{span}}},i \ne 0}^{{n_{\mathrm{span}}}} {\left( {\sum\limits_{l' = 1}^L {{\bf{g}}_{l'}^H[i]{{\bf{F}}_{\mathrm{RF}}}{{\bf{f}}_{\mathrm{BB,}l'}}} } \right)s[n - {n_{\max }} + i]} + \\ &z[n].
	\end{aligned}
\end{equation}
Then, the signal-to-interference-plus-noise ratio (SINR) of DAM with hybrid beamforming is
\begin{equation}
	{\gamma} = \frac{{{{\left| {\sum\nolimits_{l = 1}^L {{\bf{h}}_l^H{{\bf{F}}_{\mathrm{RF}}}{{\bf{f}}_{\mathrm{BB,}l}}} } \right|}^2}}}{{\sum\nolimits_{i =  - {n_{\mathrm{span}}},i \ne 0}^{{n_{\mathrm{span}}}} {{{\left| {\sum\nolimits_{l' = 1}^L {{\bf{g}}_{l'}^H[i]{{\bf{F}}_{\mathrm{RF}}}{{\bf{f}}_{\mathrm{BB,}l'}}} } \right|}^2}}  + {\sigma ^2}}}.
	\label{SNR of DAM with Hybrid}
\end{equation}

\subsection{Tap-based DAM with Hybrid Beamforming}\label{tap-DAM}
Considering the more general tap-based channel model, the multi-path delays $\tau_l$ in \eqref{channel} may be non-integer multiples of $T_{\mathrm{s}}$. This will lead to strong channel correlation on adjacent taps, so aligning all taps with relatively strong channel strengths to a single tap would make ISI-ZF beamforming unachievable. 

To address this issue, the relatively strong taps are grouped into $L^\prime$ clusters, and the adjacent taps are grouped into one cluster, which is considered to have strong channel correlations within each cluster. 
Specifically, we need to find out the relatively weak taps with power smaller than $C\mathop {\max }\limits_{q} {\|\mathbf{h}_{\mathrm{DL}}[q]\|^2}$, where $C<1$ is a certain threshold. They are considered insignificant and set to zero. 
Then, the remaining non-zero taps are grouped based on their adjacency, i.e., consecutive non-zero taps are grouped into a single cluster, with the assumption that taps within the same cluster exhibit strong channel correlations.
Consequently, the strongest tap within each cluster can be aligned to the same tap through \textit{delay pre-compensation}, and finally, minimum mean-square error (MMSE) beamforming can be applied. Similar to \eqref{x[n]p}, the transmitted signal is
\begin{equation}
	{\bf{x}}[n] = {{\bf{F}}_{\mathrm{RF}}}\sum\nolimits_{l = 1}^{L^\prime} {{{\bf{f}}_{\mathrm{BB},l}}s[n - {\kappa _l}]},
	\label{x[n]t}
\end{equation}
where $L^\prime \le L$ denotes the number of channel taps that need to be aligned.
By substituting \eqref{x[n]t} to \eqref{y[n]t}, the received signal can be expressed as
\begin{equation}
	\begin{aligned}
		{y}[n] = \sum\limits_{q = 0}^Q {{\mathbf{h}}_{\mathrm{DL}}^H[q]\sum\limits_{l = 1}^{L'} {{\bf{F}}_{\mathrm{RF}}}{{{\mathbf{f}}_{\mathrm{BB},l}}} s[n - {\kappa _l} - q] + z[n]} .
	\end{aligned}
\end{equation}
By letting ${{\bf{F}}_{\mathrm{RF}}}{{\bf{f}}_{\mathrm{BB,}l}} = {{\bf{f}}_l},\forall l$ to determine the optimal solution for fully digital beamforming design first, $y[n]$ can be further expressed as
\begin{equation}
	{y}[n] = \sum\limits_{q = 0}^Q {\sum\limits_{l = 1}^{L'} {{\mathbf{ h}}_{\mathrm{DL}}^H[q]{{\mathbf{f}}_{l}}} s[n - {\kappa _l} - q] + z[n]}.
	\label{y[n]t2}
\end{equation}

Let ${q_l},l = 1, \ldots, L^{\prime}$ be the strongest tap within the $l$th cluster and align taps $\{q_l\}_{l = 1}^{L^{\prime}}$ to the tap ${q_{\max }} = \mathop {\max }\limits_{1 \le l \le L'} {q_l}$, i.e., let ${\kappa _l} = {q_{\max }} - {q_l}$, \eqref{y[n]t2} can be further expressed as
\begin{equation}
	\begin{aligned}
		{y}[n] &= \left( {\sum\limits_{l = 1}^{L^\prime} {{\mathbf{h}}_{\mathrm{DL}}^H[q_l]{{\mathbf{f}}_{l}}} } \right)s[n - {q_{\max }}]\\
		&+ \sum\limits_{l = 1}^{L^\prime} {\sum\limits_{q \ne {q_l}}^Q {{\mathbf{h}}_{\mathrm{DL}}^H[q]{{\mathbf{f}}_{l}}s[n - {q_{\max }} + {q_l} - q]} }  + z[n].
	\end{aligned}
	\label{y[n]t3}
\end{equation}
Observing \eqref{y[n]t3}, it becomes apparent that the first term contributes to the desired signal if the receiver is locked to the tap ${q_{\max }}$, whereas the second term represents the ISI.
Since the terms with the same delay correspond to the same symbols, which need to be grouped, we define the effective channel as
\begin{equation}
	{\mathbf{g}}_l^H[i] = \left\{ {\begin{array}{*{20}{l}}
			{{\mathbf{h}}_{\mathrm{DL}}^H[q],{\rm{if }}\exists l \in \left\{ {1, \ldots, L'} \right\},\mathrm{s}\mathrm{.t}\mathrm{.}{q_l} - q = i,}\\
			{{\mathbf{0}},\mathrm{otherwise},}
	\end{array}} \right.
	\label{gl[i]}
\end{equation}
where $i \in \left\{ { \pm 1, \ldots,  \pm Q} \right\}$. Thus, \eqref{y[n]t3} can be further expressed as
\begin{equation}
	\begin{aligned}
		y[n] &= \left( {\sum\limits_{l = 1}^{L'} {{\mathbf{h}}_{\mathrm{DL}}^H[q_l]{{\bf{f}}_{l}}} } \right)s[n - {q_{\max }}]\\
		&+ \sum\limits_{i =  - Q,i \ne 0}^Q {\left( {\sum\limits_{l = 1}^{L'} {{\bf{g}}_l^H[i]{{\bf{f}}_{l}}} } \right)s[n - {q_{\max }} + i]}  + z[n].
	\end{aligned}
\end{equation}

The resulting SINR is
\begin{equation}
	\begin{aligned}
		\gamma & =\frac{\left|\sum_{l=1}^{L^{\prime}} {\mathbf{h}}_{\mathrm{DL}}^H\left[q_l\right] \mathbf{f}_{l}\right|^2}{\sum_{i=-Q, i \neq 0}^Q\left|\sum_{l=1}^{L^{\prime}} \mathbf{g}_l^H[i] \mathbf{f}_{l}\right|^2+\sigma^2} \\
		& =\frac{{\mathbf{\bar f}}^H {\mathbf{\bar h}}_{\mathrm{DL}} {\mathbf{\bar h}}_{\mathrm{DL}}^H {\mathbf{\bar f}}}{{\mathbf{\bar f}}^H\left(\sum_{i=-Q, i \neq 0}^Q {\mathbf{\bar g}}[i] {\mathbf{\bar g}}^H[i]+\sigma^2 \mathbf{I} /\left\|{\mathbf{\bar f}}\right\|^2\right) {\mathbf{\bar f}}},
	\end{aligned}
	\label{SINR}
\end{equation}
where ${\mathbf{\bar h}}_{\mathrm{DL}} = {[{\mathbf{h}}_{\mathrm{DL}}^T[q_1], \ldots ,{\mathbf{h}}_{\mathrm{DL}}^T[q_{L^\prime}]]^T} \in \mathbb{C}^{{M_{\mathrm{t}}}L' \times 1}$, ${{\mathbf{\bar f}}} = {[{\mathbf{f}}_{1}^T, \ldots ,{\mathbf{f}}_{L'}^T]^T} \in \mathbb{C}^{{M_{\mathrm{t}}}L' \times 1}$, and ${\mathbf{\bar g}}[i] = {[{\mathbf{g}}_1^T[i], \ldots ,{\mathbf{g}}_{L'}^T[i]]^T} \in \mathbb{C}^{{M_{\mathrm{t}}}L' \times 1}$. To maximize the spectral efficiency of tap-based DAM, SINR in~\eqref{SINR} should be maximized and thus the digital beamforming vector should be designed as
\begin{equation}
	{\mathbf{\bar f}}^{\mathrm{MMSE}}=\sqrt{P} \mathbf{C}^{-1} {\mathbf{\bar h}}_{\mathrm{DL}} /\left\|\mathbf{C}^{-1} {\mathbf{\bar h}}_{\mathrm{DL}}\right\|.
	\label{f_MMSE}
\end{equation}
where $\mathbf{C} \triangleq \sum_{i=-Q, i \neq 0}^Q \overline{\mathbf{g}}[i] \overline{\mathbf{g}}^H[i]+\sigma^2 / P \mathbf{I}$, with ${\left\| {{{{\mathbf{\bar f}}}}} \right\|^2} = P$. Thus, the optimal fully digital beamforming matrix based on MMSE can be obtained, noted by $\bf{F}_{\mathrm{opt}}^{\mathrm{MMSE}}$. Exploiting the idea of hybrid beamforming in the previous subsection, substituting $\bf{F}_{\mathrm{opt}}^{\mathrm{MMSE}}$ into the optimization problem \eqref{hybrid problem 2} or \eqref{hybrid problem p} and solving it, the hybrid analog/digital beamforming design can be obtained.

\section{Beam Alignment Based DAM}\label{BA-DAM}
For the aforementioned DAM technique based on fully digital or hybrid analog/digital beamforming, it is necessary to estimate the channel state information (CSI). However, the estimation overhead and complexity of the complete CSI matrix become impractical as the number of antenna elements increases. Besides, for massive MIMO systems with DAM, significant computational overhead is incurred. In the following, we propose a codebook-based beam alignment DAM technique to reduce the complexity of channel estimation.

Based on the transmitter architecture in Fig.~\ref{DAM1}, achieving beam alignment based DAM involves two stages: codebook-based analog beamforming and digital beamforming.

\subsection{Codebook-based Analog Beamforming Design}

To reduce the cost of channel estimation, codebook-based analog beamforming can be used to align the most dominant channel taps. Commonly used equivalent DFT codebook~\cite{BA+CE} is given by
\begin{equation}
	\mathcal{F}_{\mathrm{DFT}}=\left\{\mathbf{a}_{\mathrm{t}}(\hat{\theta}): \hat{\theta} \in \Theta\right\},
	\label{DFT}
\end{equation}
\begin{equation}
	\Theta  = \left\{ {\hat \theta :(1 + \sin (\hat \theta ))/2 = \frac{{{m_\mathrm{t}} - 1}}{{{M_\mathrm{t}}}},{m_\mathrm{t}} \in [1,{M_\mathrm{t}}]} \right\}.
\end{equation}

Based on the transmitter structure shown in Fig.~\ref{DAM1}, to initially determine the analog beamforming matrix ${\bf{F}}_{\mathrm{RF}}$, the $l$th column of the analog beamforming matrix denoted by ${{\bf{f}}_{\mathrm{RF},l}}{ \in \mathbb{C}^{{M_{\mathrm{t}}} \times 1}}$ is selected from the DFT codebook in \eqref{DFT}, and the digital beamforming matrix ${\bf{F}}_{\mathrm{BB}}$ is set as the identity matrix, i.e., ${{\mathbf{f}}_{\mathrm{RF},l}} \in {\mathrm{{\cal F}}_\mathrm{DFT}}$, ${{\bf{F}}_{\mathrm{BB}}} = {{\mathbf{I}}_{{M_{\mathrm{RF}}}}}$. Note that \textit{delay pre-compensation} is neither required nor achievable at this stage due to the lack of CSI. So one difference from Fig.~\ref{DAM1} is that in this stage, instead of digital beamforming the signal with \textit{delay pre-compensation}, we directly input the pilot sequences for digital beamforming. 
Thus, the transmitted signal can be expressed as
\begin{equation}
	\begin{aligned}
		\mathbf{x}[n] & =\mathbf{F}_{\mathrm{RF}} \mathbf{F}_{\mathrm{BB}} \mathbf{P m}[n]=\mathbf{F}_{\mathrm{RF}} \mathbf{P m}[n] \\
		& =\sum\nolimits_{l=1}^{M_{\mathrm{RF}}} \sqrt{p_l} \mathbf{f}_{\mathrm{RF}, l} m_l[n],
	\end{aligned}
	\label{x[n]}
\end{equation}
where ${\bf{m}}[n]={\left[ {{m_1}[n], {m_2}[n], \ldots, {m_{{M_{\mathrm{RF}}}}}[n]} \right]^T}{ \in \mathbb{C}^{{M_{\mathrm{RF}}} \times 1}}$ includes $M_\mathrm{RF}$ pilot sequences, and $\mathbf{P}=\operatorname{diag}\left(\sqrt{p_1}, \sqrt{p_2}, \ldots, \sqrt{p_{M_{\mathrm{RF}}}}\right) \in \mathbb{C}^{M_{\mathrm{RF}} \times M_{\mathrm{RF}}}$ is the power allocation matrix with ${p_l}$, $\forall l=1,\ldots, {M_{\mathrm{RF}}}$ denoting the transmit power allocated to the $l$th sequence.
Note that $m_l[n]$ is independent for different $l$ and $n$, and it has the normalized power, i.e., $\mathbb{E}\left[\left|m_l[n]\right|^2\right]=1$. The transmit power of~\eqref{x[n]} is
\begin{equation}
	\begin{aligned}
		& \mathbb{E}\left[\|\mathbf{x}[n]\|^2\right]=\sum\nolimits_{l=1}^{M_{\mathrm{RF}}} \mathbb{E}\left[\left\|\sqrt{p_l} \mathbf{f}_{\mathrm{RF}, l} m_l[n]\right\|^2\right] \\
		& =\sum\nolimits_{l=1}^{M_{\mathrm{RF}}}p_l\left\| \mathbf{f}_{\mathrm{RF}, l}\right\|^2 \leq P.
	\end{aligned}
\end{equation}
Assuming an equal allocation of transmit power, i.e., $p_l = \bar p, \forall l$, we can further obtain
\begin{equation}
	\mathbb{E}\left[\|\mathbf{x}[n]\|^2\right]=\bar p \sum\nolimits_{l=1}^{M_{\mathrm{RF}}}\left\|\mathbf{f}_{\mathrm{RF}, l}\right\|^2=\bar p\left\|\mathbf{F}_{\mathrm{RF}}\right\|_F^2 \leq P.
\end{equation}
From this, it can be derived that $\bar p \leq P /\left\|\mathbf{F}_{\mathrm{RF}}\right\|_F^2$.

Note that ${m_l}[n], \forall l \in [1,{M_{\mathrm{RF}}}]$ can be a pseudo-random sequence with period $T_m$, e.g., Zadoff-Chu (ZC) sequence, which has good autocorrelation and cross-correlation properties. These properties help the receiver to separate different beams in the codebook that are being transmitted simultaneously. Let the autocorrelation and cross-correlation of the sequence be 
\begin{equation}
	R_{m_l}[n]=\left\{\begin{array}{l}
		\frac{1}{T_m} \sum_{i=0}^{T_m-1} m_l[i] m_l^*[i]=R_{\max }, n=0, \\
		\frac{1}{T_m} \sum_{i=0}^{T_m-1} m_l[i] m_l^*[i-n] \approx 0, n \in [1,T_m-1],
	\end{array}\right.
	\label{auto-c}
\end{equation}
\begin{equation}
	{R_{{m_l},{m_{l'}}}}[n] = \frac{1}{{{T_m}}}\sum\limits_{i = 0}^{{T_m} - 1} {{m_l}[i]m_{l'}^ * [i - n]}  \approx 0, l \ne l'.
	\label{cross-c}
\end{equation}

It can be inferred from \eqref{x[n]} that since there are $M_\mathrm{RF}$ RF chains, the transmitter is able to search $M_\mathrm{RF}$ beams in the codebook simultaneously for each transmission. And there are a total of $M_\mathrm{t}$ beams in the codebook, so the transmitter needs to transmit $\left\lceil M_{\mathrm{t}} / M_{\mathrm{RF}}\right\rceil$ times. As a result, the transmit signal for the $t$th transmission can be expressed as
\begin{equation}
	\mathbf{x}_t[n]=\sqrt{\bar p} \sum\nolimits_{l=1}^{M_{\mathrm{RF}}} \mathbf{f}_{\mathrm{RF}, t, l} m_l[n], t \in\left[1,\left\lceil M_{\mathrm{t}} / M_{\mathrm{RF}}\right\rceil\right].
	\label{x_t[n]}
\end{equation}

By substituting \eqref{x_t[n]} to \eqref{y[n]t}, the received signal during the beam searching phase can be expressed as
\begin{equation}
	\begin{aligned}
		y_t[n] & =\sum_{q=0}^Q \mathbf{h}_{\mathrm{DL}}^H[q] \mathbf{x}_t[n-q]+z_t[n] \\
		& =\sqrt{\bar p}\sum_{q=0}^Q \sum_{l=1}^{M_{\mathrm{RF}}} \mathbf{h}_{\mathrm{DL}}^H[q] \mathbf{f}_{\mathrm{RF}, t, l} m_l[n-q]+z_t[n].
	\end{aligned}
	\label{y_t[n]}
\end{equation}

Because of the cross-correlation properties of pilot sequences in \eqref{cross-c},  the $M_\mathrm{RF}$ pilot sequences transmitted from the BS can be approximately separated at the UE by passing the received signal in \eqref{y_t[n]} through a bank of matched filters where the $l$th filter has impulse response $m_l^ * [ - n]$. Thus, it can be obtained that
\begin{equation}
	\hspace{-2ex}
	\begin{aligned}
		y_{t, l}[n] & =\frac{1}{T_m} \sum_{i=Q}^{Q+T_m-1} y_t[i] m_l^*[i-n] \\
		& =\frac{\sqrt{\bar p}}{T_m} \sum_{i=Q}^{Q+T_m-1} \sum_{q=0}^Q \sum_{l^{\prime}=1}^{M_{\mathrm{RF}}} \mathbf{h}_{\mathrm{DL}}^H[q] \mathbf{f}_{\mathrm{RF}, t, l^{\prime}} m_{l^{\prime}}[i-q] m_l^*[i-n]\\
		&+\frac{1}{T_m} \sum_{i=Q}^{Q+T_m-1} z_t[i] m_l^*[i-n] \\
		& =\frac{\sqrt{\bar p}}{T_m} \sum_{q=0}^Q \sum_{l^{\prime}=1}^{M_{\mathrm{RF}}} \mathbf{h}_{\mathrm{DL}}^H[q] \mathbf{f}_{\mathrm{RF}, t, l^{\prime}} \sum_{i=Q}^{Q+T_m-1} m_{l^{\prime}}[i-q] m_l^*[i-n] \\
		&+z_t^c[n] \\
		& =\sqrt{\bar p}\sum_{q=0}^Q \sum_{l^{\prime}=1}^{M_{\mathrm{RF}}} \mathbf{h}_{\mathrm{DL}}^H[q] \mathbf{f}_{\mathrm{RF}, t, l^{\prime}} R_{m_{l^\prime}, m_l}[n-q]+z_t^c[n], \\
		 n &\in [0,T_m-1].
	\end{aligned}
	\label{y_t,l[n]}
\end{equation}

Note that the good cross-correlation properties of sequence ${m_l}[n], \forall l \in [1,{M_{\mathrm{RF}}}]$ are based on a complete period $T_m$. Therefore, in \eqref{y_t,l[n]}, when performing the correlation operation, the range of $i$ is considered to be $[Q,Q + {T_m} - 1]$, implying that $0 \le i - q \le Q + {T_m} - 1$. Since the period of ${R_{{m_l},{m_{l'}}}}[n]$ is $T_m$, it follows that the period of ${y_{t,l}}[n]$ is also $T_m$, so we take the range of $n$ in \eqref{y_t,l[n]} to be $[0,T_m-1]$.
To ensure that the correlation operation at the receiver is based on a complete period, the transmitter should send the pilot sequences periodically. Assuming that the length of the transmitted sequence in \eqref{x_t[n]} is $a{T_m}$, since the index of ${m_{l'}}[i - q]$ in \eqref{y_t,l[n]} is in the range of $0 \le i - q \le Q + {T_m} - 1$, it should satisfy $a{T_m} - 1 \ge Q + {T_m} - 1$. Therefore, it is necessary to transmit at least $a = {Q \mathord{\left/{\vphantom {Q {{T_m} + 1}}} \right.	\kern-\nulldelimiterspace} {{T_m} + 1}}$ periods of the pilot sequences, so that $a{T_m} \ge Q + {T_m} $, where $T_m$ needs to be large enough to ensure good cross-correlation properties of the pilot sequences. Note that, when choosing the period of pilot sequences, there is a tradeoff between maintaining good cross-correlation properties and avoiding a significant increase in beam search time.

With~\eqref{cross-c}, \eqref{y_t,l[n]} can be further simplified as
\begin{equation}
	\hspace{-1ex}
	{y_{t,l}}[n] \approx \sqrt{\bar p}\sum\limits_{q = 0}^Q {{\mathbf{h}}_{\mathrm{DL}}^H[q]{{\mathbf{f}}_{\mathrm{RF},t,l}}{R_{{m_l}}}[n - q]}  + z_t^c[n], n \in [0,T_m-1].
	\label{y_t,l[n]2}
\end{equation}
Note that this approximation comes from the good cross-correlation properties in~\eqref{cross-c} of the pseudo-random sequence, and the accuracy of the approximation can be affected by the length of the sequences as well as the specific design scheme.

From \eqref{auto-c}, it is known that when the beam ${{\mathbf{f}}_{\mathrm{RF},t,l}}$ aligns with the strong channel tap of ${{\mathbf{h}}_{\mathrm{DL}}^H}[q]$ and satisfies $q=n+b T_m, \forall b \in \mathbb{Z}$, ${y_{t,l}}[n]$ in \eqref{y_t,l[n]2} can achieve a large value relatively. To determine which beam can align with the strong channel tap and decrease the impact of noise, calculating the power after summing ${y_{t,l}}[n]$, we can obtain
\begin{equation}
	\hspace{-3ex}
	\begin{aligned}
		{r_{t,l}} &= {\left| {\sum\limits_{n = 0}^{{T_m} - 1} {{y_{t,l}}[n]} } \right|^2}\\
		&= \bar p {\left| {\sum\limits_{n = 0}^{{T_m} - 1} {\left( {\sum\limits_{q = 0}^Q {{\mathbf{h}}_{\mathrm{DL}}^H[q]{{\mathbf{f}}_{\mathrm{RF},t,l}}{R_{{m_l}}}[n - q]}  + z_t^c[n]} \right)} } \right|^2}.
	\end{aligned}
	\label{r_t,l}
\end{equation}

As a result, by searching all ${M_\mathrm{t}}$ beams in the codebook and sorting $r_{t,l}, \forall{t,l}$ in descending order, ${M_{\mathrm{RF}}}$ vectors ${{\mathbf{f}}_{\mathrm{RF},t,l}}$ corresponding to the first ${M_{\mathrm{RF}}}$ largest $r_{t,l}$ will be selected, denoted as $\{ {{\mathbf{f}}_{\mathrm{RF},l}^{\mathrm{opt}}} \}_{l = 1}^{{M_{\mathrm{RF}}}}$, which are used to form the analog beamforming matrix $\mathbf{F}_{\mathrm{RF}}^\mathrm{opt}=\left[\mathbf{f}_{\mathrm{RF}, 1}^{\mathrm{opt}}, \ldots \mathbf{f}_{\mathrm{RF}, M_{\mathrm{RF}}}^{\mathrm {opt }}\right] \in \mathbb{C}^{M_{\mathrm{t}} \times M_{\mathrm{RF}}}$.

The above codebook-based analog beamforming design for DAM with hybrid beamforming is summarized in Algorithm~\ref{algorithm 1}.
\begin{algorithm}[t!]
	\caption{Codebook-based Analog Beamforming Design}
	\label{algorithm 1}	
	\LinesNumbered 
	\SetAlgoLined
	\KwIn{Transmit power $P$, DFT codebook $\mathcal{F}_{\mathrm{DFT}}$, and pilot sequences $\{m_l[n]\}^{M_{\mathrm{RF}}}_{l=1}$.}
	\KwOut{The optimal analog beamforming matrix $\mathbf{F}_{\mathrm{RF}}^\mathrm{opt}$.}
	\For{$t=1:\left\lceil M_{\mathrm{t}} / M_{\mathrm{RF}}\right\rceil$}{
		Use $\mathbf{f}_{\mathrm{RF}, t, l} \in \mathcal{F}_{\mathrm{DFT}}, \forall l \in [1,M_{\mathrm{RF}}]$ as analog beamforming vector and transmit the signal in~\eqref{x_t[n]}\;
		Pass the received signal in~\eqref{y_t[n]} through a bank of matched filters to obtain $y_{t, l}[n]$ in~\eqref{y_t,l[n]}\;
		Calculate the power $r_{t,l}$ in~\eqref{r_t,l}\;
	}
	Sort $r_{t,l}, \forall t,l$ in descending order and select ${M_{\mathrm{RF}}}$ vectors ${{\mathbf{f}}_{\mathrm{RF},t,l}}$ corresponding to the first ${M_{\mathrm{RF}}}$ largest $r_{t,l}$ to form the optimal beamforming matrix $\mathbf{F}_{\mathrm{RF}}^\mathrm{opt}$\;
\end{algorithm}
\subsection{Digital Beamforming Design}
After determining the analog beamforming matrix, the digital beamforming matrix needs to be further determined, which can be based on the tap-based DAM method introduced earlier. Note that the analog beamforming matrix is already determined here, so this stage is equivalent to the fully digital beamforming design, and the transmitted signal is given by
\begin{equation}
	{\mathbf{x}}[n] = {{\mathbf{F}}_{\mathrm{RF}}^{\mathrm{opt}}}\sum\nolimits_{l = 1}^{L'} {{{\mathbf{f}}_{\mathrm{BB},l}}s[n - {\kappa _l}]}.
	\label{x[n]2}
\end{equation}
By substituting \eqref{x[n]2} to \eqref{y[n]t}, the received signal is given by
\begin{equation}
	\begin{aligned}
		{y}[n] &= \sum\limits_{q = 0}^Q {{\mathbf{h}}_{\mathrm{DL}}^H[q]{\mathbf{x}}[n - q] + z[n]} \\
		&= \sum\limits_{q = 0}^Q {{\mathbf{h}}_{\mathrm{DL}}^H[q]{{\bf{F}}_{\mathrm{RF}}^{\mathrm{opt}}}\sum\limits_{l = 1}^{L'} {{{\mathbf{f}}_{\mathrm{BB},l}}} s[n - {\kappa _l} - q] + z[n]} .
	\end{aligned}
\end{equation}
By letting the downlink equivalent channel tap ${{\mathbf{\tilde h}}_{\mathrm{DL}}}^H[q] = {{\mathbf{h}}_{\mathrm{DL}}^H}[q]{\mathbf{F}}_{\mathrm{RF}}^{\mathrm{opt}} \in \mathbb{C}^{1 \times {M_\mathrm{RF}}}$, $y[n]$ can be further expressed as
\begin{equation}
	{y}[n] = \sum\limits_{q = 0}^Q {\sum\limits_{l = 1}^{L'} {{\mathbf{\tilde h}}_{\mathrm{DL}}^H[q]{{\mathbf{f}}_{\mathrm{BB},l}}} s[n - {\kappa _l} - q] + z[n]}.
	\label{y[n]2}
\end{equation}
Obviously, it is of the same form as \eqref{y[n]t2}, the only difference is that the equivalent channel by considering the effect of analog beamforming is used here, which has a much smaller dimension ${M_{\mathrm{RF}}} \ll {M_\mathrm{t}}$. Therefore, the complexity of channel estimation can be effectively reduced. Assume that the equivalent channel is estimated, with ${\mathbf{g}}_l^H[i]$ defined in \eqref{gl[i]}, the resulting SINR is given by
\begin{equation}
	\gamma =\frac{\left|\sum_{l=1}^{L^{\prime}} {\mathbf{\tilde h}}_{\mathrm{DL}}^H\left[q_l\right] \mathbf{f}_{\mathrm{BB}, l}\right|^2}{\sum_{i=-Q, i \neq 0}^Q\left|\sum_{l=1}^{L^{\prime}} \mathbf{g}_l^H[i] \mathbf{f}_{\mathrm{BB}, l}\right|^2+\sigma^2}.
	\label{SINR_BA1}
\end{equation} 
To maximize the spectral efficiency, it is equivalent to maximizing the SINR in~\eqref{SINR_BA1}, which can be written as
\begin{equation}
	\begin{aligned}
		\mathop {\max}\limits_{\{{\bf{f}}_{\mathrm{BB},l}\}^{L^\prime}_{l=1}} \enspace&\frac{\left|\sum_{l=1}^{L^{\prime}} {\mathbf{\tilde h}}_{\mathrm{DL}}^H\left[q_l\right] \mathbf{f}_{\mathrm{BB}, l}\right|^2}{\sum_{i=-Q, i \neq 0}^Q\left|\sum_{l=1}^{L^{\prime}} \mathbf{g}_l^H[i] \mathbf{f}_{\mathrm{BB}, l}\right|^2+\sigma^2}\\
		\text{s.t.}\enspace&{\left\| {{{\bf{F}}_{\mathrm{RF}}^{\mathrm{opt}}}{{\bf{F}}_{\mathrm{BB}}}} \right\|_F^2} =\sum\limits_{l = 1}^L {\left\| {{{\bf{F}}_{\mathrm{RF}}^{\mathrm{opt}}}{{\bf{f}}_{\mathrm{BB},l}}} \right\|^2} \le P.
	\end{aligned}
	\label{P_MMSE1}
\end{equation}
Note that the power constraints of this optimization problem include ${{\bf{F}}_{\mathrm{RF}}^{\mathrm{opt}}}$ which is determined, not just only ${{\bf{f}}_{\mathrm{BB},l}}$, making the solution in~\eqref{f_MMSE} inapplicable here. To this end, let $
{\bf{F}}_{\mathrm{RF}}^{\mathrm{opt}}{{\mathbf{f}}_{\mathrm{BB},l}} = {{\mathbf{v}}_l}$. From the previous section, it can be seen that the columns $\{ {\mathbf{f}}_{\mathrm{RF},l}^{\mathrm{opt}} \}_{l = 1}^{{M_{\mathrm{RF}}}}$ that make up ${{\bf{F}}_{\mathrm{RF}}^{\mathrm{opt}}}$ are different, each corresponding to a different AoD. Therefore, they are linearly independent, meaning ${\bf{F}}_{\mathrm{RF}}^{\mathrm{opt}}$ is of full column rank. Thus, the digital beamforming vector can be expressed as $
{{\mathbf{f}}_{\mathrm{BB},l}} = {({\bf{F}}_{\mathrm{RF}}^{\mathrm{opt}})^{\dagger}}{{\mathbf{v}}_l}, \forall l$, where $(\mathbf{F}_{\mathrm{RF}}^{\mathrm{opt}})^{\dagger}$ is the pseudo-inverse of ${\bf{F}}_{\mathrm{RF}}^{\mathrm{opt}}$. Then, the SINR in~\eqref{SINR_BA1} can be further expressed as
\begin{equation}
	\begin{aligned}
		\gamma &= \frac{\left|\sum_{l=1}^{L'}{\mathbf{\tilde h}}_{\mathrm{DL}}^{H}\left[q_{l}\right](\mathbf{F}_{\mathrm{RF}}^{\mathrm{opt}})^{\dagger} {{\mathbf{v}}_l}\right|^{2}}{\sum_{i=-Q, i \neq 0}^{Q}\left|\sum_{l=1}^{L'} \mathbf{g}_{l}^{H}\left[i\right](\mathbf{F}_{\mathrm{RF}}^{\mathrm{opt}})^{\dagger} {{\mathbf{v}}_l}\right|^{2} + \sigma^{2}} \\
		&= \frac{{\mathbf{\bar v}}^{H} {\mathbf{\bar h}}_{\mathrm{DL}} {{\mathbf{\bar h}}}_{\mathrm{DL}}^{H} {\mathbf{\bar v}}}{{\mathbf{\bar v}}^{H} \left( \sum_{i=-Q, i \neq 0}^{Q} {\mathbf{\bar g}}[i] \bar{\mathbf{g}}^{H}[i] + \sigma^{2} \mathbf{I} / \|{\mathbf{\bar v}}\|^{2} \right) {\mathbf{\bar v}}},
	\end{aligned}
\end{equation}
where ${\mathbf{\bar h}}_{\mathrm{DL}} = {[{\mathbf{\tilde h}}_{\mathrm{DL}}^T[q_1]((\mathbf{F}_{\mathrm{RF}}^{\mathrm{opt}})^{\dagger})^{*}, \ldots ,{\mathbf{\tilde h}}_{\mathrm{DL}}^T[q_{L^\prime}]((\mathbf{F}_{\mathrm{RF}}^{\mathrm{opt}})^{\dagger})^{*}]^T} \in \mathbb{C}^{{M_{\mathrm{t}}}L' \times 1}$, ${{\mathbf{\bar v}}} = {[{\mathbf{v}}_{1}^T, \ldots ,{\mathbf{v}}_{L'}^T]^T} \in \mathbb{C}^{{M_{\mathrm{t}}}L' \times 1}$, and ${\mathbf{\bar g}}[i] = {[{\mathbf{g}}_1^T[i]((\mathbf{F}_{\mathrm{RF}}^{\mathrm{opt}})^{\dagger})^{*}, \ldots ,{\mathbf{g}}_{L'}^T[i]((\mathbf{F}_{\mathrm{RF}}^{\mathrm{opt}})^{\dagger})^{*}]^T} \in \mathbb{C}^{{M_{\mathrm{t}}}L' \times 1}$. Thus, the problem in~\eqref{P_MMSE1} is equivalent to
\begin{equation}
	\begin{aligned}
		\mathop {\max}\limits_{{\mathbf{\bar v}}} \enspace&\frac{{\mathbf{\bar v}}^{H} {\mathbf{\bar h}}_{\mathrm{DL}} {{\mathbf{\bar h}}}_{\mathrm{DL}}^{H} {\mathbf{\bar v}}}{{\mathbf{\bar v}}^{H} \left( \sum_{i=-Q, i \neq 0}^{Q} {\mathbf{\bar g}}[i] \bar{\mathbf{g}}^{H}[i] + \sigma^{2} \mathbf{I} / \|{\mathbf{\bar v}}\|^{2} \right) {\mathbf{\bar v}}}\\
		\text{s.t.}\enspace&\|{\mathbf{\bar v}}\|^{2} = P.
	\end{aligned}
	\label{P_MMSE2}
\end{equation}
It is noted that the objective function is a generalized Rayleigh quotient with respect to ${\mathbf{\bar v}}$, which is maximized by the MMSE beamforming 	${\mathbf{\bar v}}^{\mathrm{MMSE}}=\sqrt{P} \mathbf{C}^{-1} {\mathbf{\bar h}}_{\mathrm{DL}} /\left\|\mathbf{C}^{-1} {\mathbf{\bar h}}_{\mathrm{DL}}\right\|$,
where $\mathbf{C} \triangleq \sum_{i=-Q, i \neq 0}^Q {\mathbf{\bar g}}[i] {\mathbf{\bar g}}^H[i]+\sigma^2 / P \mathbf{I}$. Therefore, the optimal digital beamforming vector can be obtained as ${\mathbf{f}}{_{\mathrm{BB},l}^{\mathrm{MMSE}}} = {({\bf{F}}_{\mathrm{RF}}^{\mathrm{opt}})^{\dagger} }{\mathbf{v}}_l^{\mathrm{MMSE}}$.

\section{DAM-OFDM with Hybrid Beamforming} \label{HB-DAM-OFDM}
\begin{figure*}[t!]
	\centerline{\includegraphics[width=\textwidth]{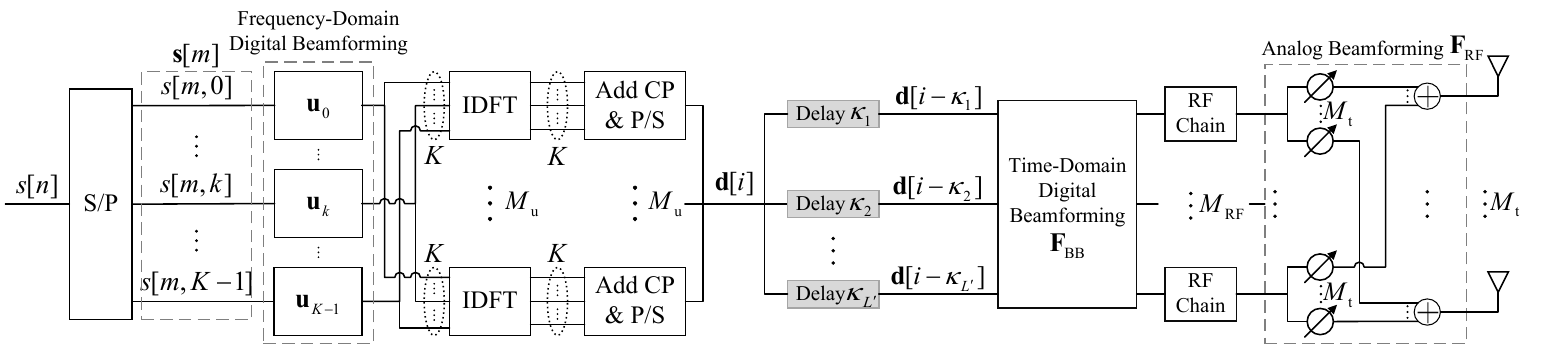}}
	\captionsetup{justification=raggedright,singlelinecheck=false}
	\caption{Transmitter architecture of the DAM-OFDM communication system with hybrid beamforming.}
	\label{DAM-OFDM}
\end{figure*}
As discussed in previous sections, for channels with fractional delays, since adjacent channel taps in one cluster have strong correlations, DAM may not completely eliminate ISI. The work in~\cite{DAM-OFDM} proposed DAM-OFDM, which integrates DAM to flexibly manipulate the channel delay spread for more efficient OFDM transmission. Inspired by this, it is possible to align each cluster into one cluster using DAM, thereby reducing channel delay spread, and further overcoming the residual ISI with OFDM. To reduce hardware cost and power consumption, hybrid analog/digital beamforming will also be used for DAM-OFDM in the following.

Based on the transmitter architecture of DAM-OFDM with hybrid beamforming in Fig.~\ref{DAM-OFDM}, after OFDM processing, the $n$th sample of the $m$th OFDM symbol in the time-domain~ is given by\cite{DAM-OFDM}
\begin{equation}
	\mathbf{\bar{x}}_\mathrm{t}[m,n] = \frac{1}{\sqrt{K}} \sum_{k=0}^{K-1} \mathbf{u}_k s[m,k] e^{j \frac{ 2 \pi}{K} k n}, n = -N_{\text{CP}}, \ldots, K-1,
\end{equation}
where $K$ is the number of sub-carriers used in OFDM, $\mathbf{u}_k \in \mathbb{C}^{{M_{\mathrm{u}}} \times 1 }$ denotes the digital baseband beamforming vector for sub-carrier $k$, $s[m,k]$ denotes the information symbol carried by the $k$th sub-carrier in the $m$th OFDM symbol, and $N_{\text{CP}}$ denotes the length of CP.
With parallel to serial conversion, the samples of all OFDM symbols $\mathbf{\bar{x}}_t[m,n]$ are concatenated as a time-domain sequence, which is given by
\begin{equation}
	\mathbf{d}[i] \triangleq \mathbf{\bar{x}}_\mathrm{t}[m,n], \quad \forall n \in [-N_{\text{CP}}, K-1],
\end{equation}
where  $i = m (K + N_{\text{CP}}) + n$. After DAM processing with hybrid beamforming, the transmitted signal is given by
\begin{equation}
	{\mathbf{\bar d}}[i] = {{\bf{F}}_{\mathrm{RF}}}\sum\nolimits_{l = 1}^{L'} {{{\bf{F}}_{\mathrm{BB},l}}{\bf{d}}\left[ {i - {\kappa _l}} \right]}.
\end{equation}
Based on the idea of hybrid beamforming in the previous sections, it is required to design the fully digital beamforming matrix first. Hence we let ${{\bf{F}}_{l}}={{\bf{F}}_{\mathrm{RF}}}{{\bf{F}}_{\mathrm{BB},l}} \in \mathbb{C}^{{M_\mathrm{t}} \times {M_\mathrm{u}}}$.
Thus, the transmitted signal can be further given by
\begin{equation}
	{\mathbf{\bar d}}[i] = \sum\nolimits_{l = 1}^{L'} {{{\bf{F}}_{l}}{\bf{d}}\left[ {i - {\kappa _l}} \right]}.
	\label{d[i]}
\end{equation}
It follows from~\cite{DAM-OFDM} that the transmit power of ${\mathbf{\bar d}}[i]$ in \eqref{d[i]} is
\begin{equation}
	\mathbb{E} \left[ \left\| {\mathbf{\bar d}}[i] \right\|^2 \right] = \frac{1}{K} \sum_{k=0}^{K-1} \left\| \sum_{l=1}^{L'} \mathbf{F}_l \mathbf{u}_k e^{-j \frac{2 \pi}{K} k \kappa_l} \right\|^2.
\end{equation}

Because adjacent taps may exhibit strong correlations under fractional delay conditions, according to subsection~\ref{tap-DAM}, channel taps are grouped into  $L^\prime$ clusters. The taps in each cluster are considered to have strong channel correlation. Let ${q_l},l = 1, \ldots, L^{\prime}$ be the strongest tap within the $l$th cluster and align them to the tap ${q_{\max }} = \mathop {\max }\limits_{1 \le l \le L'} {q_l}$, i.e., let ${\kappa _l} = {q_{\max }} - {q_l}$. 
Note that the MMSE beamforming will not be used here, as we aim to align all clusters to the cluster with the maximum delay ${q_{\max }}$ with the cluster-based ZF beamforming.
Let $\bar{\mathcal{Q}}_l$ denotes the set of taps contained in the $l$th cluster, where $l \in [1,L^\prime]$. Then, for the $l$th cluster, the set of taps contained in other clusters is denoted by $\mathcal{Q}_l$. If the absolute value of the maximum difference between the taps and $q_l$ in the $l$th cluster is $\bar n_{l,\mathrm{span}} \buildrel \Delta \over = \mathop {\max }\limits_{q \in \bar{\mathcal{Q}}_{l}} {\left| q-q_l \right|}$, let ${\bar n_{\mathrm{span}}} \buildrel \Delta \over = \mathop {\max }\limits_{1 \le l \le L^\prime} {\bar n_{l,\mathrm{span}}}$. Thus, the channel delay spread after the alignment of the clusters is around but not larger than $n^\prime_\mathrm{span} = 2\bar n_\mathrm{span}$, which is generally no larger than the original channel delay spread $n_\mathrm{span}$, i.e., $n^\prime_\mathrm{span} \le n_\mathrm{span}$.

Through the tap-based channel in \eqref{y[n]t}, the received signal is
\begin{equation}
	\begin{split}
		y[i] &= \sum_{q=0}^{Q} \mathbf{h}_{\mathrm{DL}}^H[q] {\mathbf{\bar d}}[i - q] + z[i] \\
		&= \sum_{q=0}^{Q} \sum_{l=1}^{L'} \mathbf{h}_{\mathrm{DL}}^H[q] \mathbf{F}_{l} \mathbf{d}[i - q - \kappa_{l}] + z[i].
	\end{split}
\end{equation}
Suppose that the strength of the channel taps that are not categorized as clusters are approximated as zero, and let ${\kappa _l} = {q_{\max }} - {q_l}$, $y[i]$ can be further expressed as
\begin{equation}
	\begin{aligned}
		y[i] = &\sum_{l=1}^{L'} \sum_{q \in \bar{\mathcal{Q}}_{l}} \mathbf{h}_{\mathrm{DL}}^H[q] \mathbf{F}_{l} \mathbf{d}[i - {q_{\max }} + ({q_l} - q)]\\
		 &+ \sum_{l=1}^{L'} \sum_{q \in {\mathcal{Q}}_{l}} \mathbf{h}_{\mathrm{DL}}^H[q] \mathbf{F}_{l} \mathbf{d}[i - {q_{\max }} + ({q_l} - q)] + z[i].
	\end{aligned}
	\label{y[i]2}
\end{equation}
Since in the first term of \eqref{y[i]2}, $q \in \bar{\mathcal{Q}}_{l}$, and thus $({q_l} - q) \in [-n^\prime_\text{span}/2, n^\prime_\text{span}/2]$, the first term represents the taps that is aligned to the cluster containing $q_{\max}$, and the second term represents the term that needs to be forced to zero. Define ${\mathbf{\bar H}}_{l} \in \mathbb{C}^{M_t \times |\mathcal{Q}_{l}|} \triangleq [\mathbf{h}_{\mathrm{DL}}[q]]_{q \in \mathcal{Q}_{l}}$. To
design $\mathbf{F}_{l}$ so that the second term in \eqref{y[i]2} can be eliminated, we should have
\begin{equation}
	{\mathbf{\bar H}}_{l}^H \mathbf{F}_{l} = \mathbf{0}_{|\mathcal{Q}_{l}| \times M_\mathrm{u}}.
\end{equation}
This ZF conditions are feasible as long as ${\mathbf{\bar H}}_{l}^H$ has a non-empty null-space~\cite{DAM-OFDM}. 
In this case, the beamforming matrix can be expressed as $\mathbf{F}_{l} = \mathbf{\bar{H}}_{l}^{\perp} \mathbf{\bar{X}}_{l}
$, where $\mathbf{\bar{H}}_{l}^{\perp} \in \mathbb{C}^{M_\text{t} \times \bar r_{l}}$ denotes an orthonormal basis for the orthogonal complement of ${\mathbf{\bar H}}_{l}$, and $\mathbf{\bar{X}}_{l} \in \mathbb{C}^{\bar r_{l} \times M_\mathrm{u}}$ is the new beamforming matrix to be designed, with $\bar r_{l} = \mathrm{rank}(\mathbf{\bar{H}}_{l}^{\perp})$. Thus, the received signal in \eqref{y[i]2} is
\begin{equation}
		y[i] = \sum_{l=1}^{L'} \sum_{q \in \bar{\mathcal{Q}}_{l}} \mathbf{h}_{\mathrm{DL}}^H[q] \mathbf{\bar{H}}_{l}^{\perp} \mathbf{\bar{X}}_{l} \mathbf{d}[i - {q_{\max }} + ({q_l} - q)] + z[i].
	\label{y[i]3}
\end{equation}

To show the impact of the resulting delay spread more clearly, those components with common delays should be grouped. To this end, for any $t \in [-n^\prime_\mathrm{span}/2, n^\prime_\mathrm{span}/2]$, define the following effective channel vector
\begin{equation}
	\mathbf{g}_{l}^H[t] = \begin{cases} 
		\mathbf{h}_{\mathrm{DL}}^H[q], & \text{if } \exists q \in \bar{\mathcal{Q}}_{l}, \text{ s.t. } q + \kappa_{l} = t + q_{\max}, \\
		0, & \text{otherwise.}
	\end{cases}
\end{equation}
Thus, \eqref{y[i]3} can be equivalently written as~\cite{DAM-OFDM}
\begin{equation}
	y[i] = \sum_{t=-n'_{\text{span}}/2}^{n'_{\text{span}}/2} \left( \sum_{l=1}^{L'} \mathbf{g}_{l}^H[t] \bar{\mathbf{H}}_{l}^{\perp} \bar{\mathbf{X}}_{l} \right) \mathbf{d}[i - t] + z[i].
	\label{y[i]ofdm}
\end{equation}
It is observed from~\eqref{y[i]ofdm} the signal $\mathbf{d}[i]$ would see an effective channel with channel delay spread $n^\prime_\mathrm{span} \le n_\mathrm{span}$, after the DAM processing. 
It is known from~\cite{DAM-OFDM} that, for DAM-OFDM with $N_\text{CP} \ge n^\prime_\mathrm{span}$, after removing the CP from the received signal in \eqref{y[i]3} and performing serial to parallel conversion, it can be further obtained
\begin{equation}
	\begin{aligned}
		y[m,n] = & \frac{1}{\sqrt{K}} \sum_{k=0}^{K-1} \left( \sum_{t=-n'_{\text{span}}/2}^{n'_{\text{span}}/2} \left( \sum_{l=1}^{L'} \mathbf{g}_{l}^H[t] \mathbf{\bar H}_{l}^{\perp} {\mathbf{\bar X}}_{l} \right) e^{-j \frac{2\pi}{K} kt} \right) \\
		& \times \mathbf{u}_k s[m,k] e^{j \frac{2\pi}{K} kn} + z[m,n], \forall n \in [0,K-1].
	\end{aligned}
	\label{y[m,n]}
\end{equation}
Let ${\mathbf{\tilde h}}^H[k]$ denotes the equivalent frequency-domain channel of the $k$th sub-carrier for DAM-OFDM, which is given by
\begin{equation}
	{\mathbf{\tilde h}}^H[k] = \frac{1}{\sqrt{K}} \sum_{t=-n'_{\text{span}}/2}^{n'_{\text{span}}/2} \left( \sum_{l=1}^{L'} \mathbf{g}_{l}^H[t] {\mathbf{\bar H}}_{l}^{\perp} {\mathbf{\bar X}}_{l} \right) e^{-j \frac{2\pi}{K} kt},  \forall k,
\end{equation}
where the coefficient $1/\sqrt{K}$ is introduced to ensure the conservation of energy between DFT and IDFT. Thus, the received signal in the frequency-domain can be obtained by taking the DFT to \eqref{y[m,n]}, which is given by
\begin{equation}
	y_\mathrm{f}[m,k] = \sqrt{K} {\mathbf{\tilde h}}^H[k] \mathbf{u}_k s[m,k] + z[m,k], \forall k \in [0, K-1].
\end{equation}
Therefore, the received SNR for sub-carrier $k$ is
\begin{equation}
	\begin{split}
		\gamma_k &= K\left| {\mathbf{\tilde h}}^H[k] \mathbf{u}_k \right|^2/\sigma^2 \\
		&= \frac{\left| \sum_{t=-n'_{\text{span}}/2}^{n'_{\text{span}}/2} \left( \sum_{l=1}^{L'} \mathbf{g}_{l}^H[t] {\mathbf{\bar H}}_{l}^{\perp} {\mathbf{\bar X}}_{l} \right) e^{-j \frac{2\pi}{K} kt} \mathbf{u}_k \right|^2}{\sigma^2}.
	\end{split}
\end{equation}
As a result, the spectral efficiency of DAM-OFDM can be maximized by jointly optimizing the time-domain beamforming matrices $\{{\mathbf{\bar X}}_{l}\}^{L^\prime}_{l=1}$ and the frequency-domain beamforming vectors $\{\mathbf{u}_k\}^{K-1}_{k=0}$ with the dimension $M_\mathrm{u}$. The optimization problem can be formulated as
\begin{equation}
	\begin{aligned}
		\max_{\{{\mathbf{\bar X}}_{l}\}_{l=1}^{L'}, \{\mathbf{u}_k\}_{k=0}^{K-1}, M_\mathrm{u}} &\frac{1}{K} \sum\nolimits_{k=0}^{K-1} \log_2 (1 + \gamma_k) \\
		\text{s.t.} \quad &\sum_{k=0}^{K-1} \left\| \sum_{l=1}^{L'} {\mathbf{\bar H}}_{l}^{\perp} {\mathbf{\bar X}}_{l} \mathbf{u}_k e^{-j \frac{2 \pi}{K} k \kappa_{l}} \right\|^2 \leq KP.
	\end{aligned}
\end{equation}
This problem can be solved using the methods outlined in steps 3 to 5 of Algorithm 1 in~\cite{DAM-OFDM}. Note that $M_\mathrm{u}$ can be chosen flexibly to obtain the optimal solution, which is denoted by $\{{\mathbf{\bar X}}^{\mathrm{opt}}_{l}\}^{L^\prime}_{l=1}$ and $\{\mathbf{u}^{\mathrm{opt}}_k\}^{K-1}_{k=0}$.
Finally, to implement DAM-OFDM based on hybrid analog/digital beamforming, according to the method in Section~\ref{HB-DAM}, analog beamforming matrix ${{\bf{F}}_{\mathrm{RF}}}$ and digital beamforming matrix ${{\bf{F}}_{\mathrm{BB},l}}$ need to be designed so that ${{\bf{F}}_{\mathrm{RF}}}{{\bf{F}}_{\mathrm{BB},l}}$ is as close to ${{\bf{F}}_{l}}={\mathbf{\bar H}}_{l}^{\perp}{\mathbf{\bar X}}^{\text{opt}}_{l}$ as possible, which can be achieved using the OMP algorithm.

\section{Simulation Results} \label{Simulation}

\begin{table}[!t]
	\renewcommand{\arraystretch}{1.3}
	\caption{Parameter settings}
	\centering
	\label{Parameter}
	\resizebox{\columnwidth}{!}{
		\begin{tabular}{l l}
			\hline\hline \\[-4mm]
			\multicolumn{1}{l}{Parameter} & \multicolumn{1}{l}{value} \\[0.5ex] \hline
			Number of RF chains & ${M_\mathrm{RF}} = 4$ \\
			Carrier frequency & $f = 28\mathrm{GHz}$ \\
			Total bandwidth & $B = 128\mathrm{MHz}$ \\
			Noise power spectral density & ${N_0} =  - 174\mathrm{dBm/Hz}$\\
			Inter-element spacing of the ULA & $d = {\lambda  \mathord{\left/{\vphantom {\lambda 2}} \right.\kern-\nulldelimiterspace} 2}$\\
			Channel coherence time & ${T_c} = 1\mathrm{ms}$\\
			Number of temporal-resolvable multi-paths & $L = 4$\\
			Delay & ${n_l} \sim \mathrm{U}[0,{\tau _{\max }}],{\tau _{\max }} = 312.5\mathrm{ns}$\\
			Number of sub-paths & ${\mu _l} \sim \mathrm{U}[0,{\mu _{\max }}],{\mu _{\max }} = 3$\\
			AoDs & ${\theta _{li}} \sim \mathrm{U}[ - 60^\circ ,60^\circ ],\forall l,i$\\
			Number of sub-carriers used in OFDM & $K = 256$\\
			Threshold for selecting significant taps & $C = 0.01$\\
			\hline\hline
		\end{tabular}
	}
\end{table}
In this section, simulation results are provided to verify the effectiveness of the proposed hybrid beamforming designs for DAM.
Unless otherwise stated, the parameter settings of the simulation results are summarized in Table \ref{Parameter}. 
Furthermore, the coefficient ${\alpha _l},\forall l$ in \eqref{hl} are generated using the model developed in~\cite{Sparsity-1}.

\begin{figure}[t!]
	\centerline{\includegraphics[width=6.5cm]{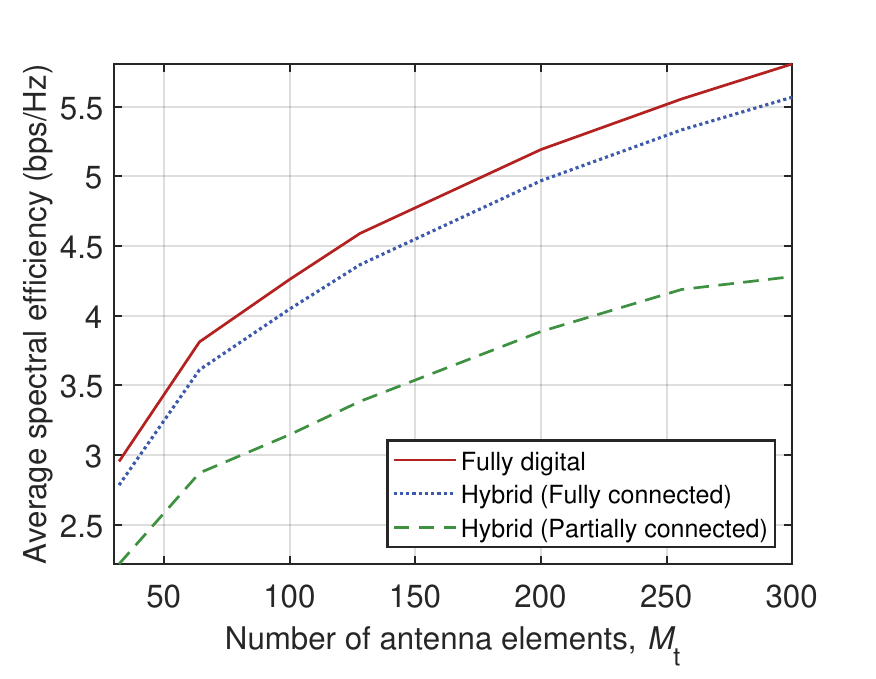}}
	\caption{Spectral efficiency of DAM for channels with integer delays.}
	\label{Rb_DAM-ID}
\end{figure}

For transmit power of $P = 30$ dBm, Fig.~\ref{Rb_DAM-ID} illustrates the average spectral efficiency versus the number of antenna elements $M_\mathrm{t}$ for DAM based on fully digital and hybrid beamforming. 
It can be observed that when the number of RF chains is $4$, which is equal to that of multi-paths, 
the average spectral efficiency of DAM based on fully digital beamforming and that of DAM based on hybrid beamforming with fully connected structure exhibit an increasing trend with similar rates of growth as the number of transmit antennas increases. This is because the ability of hybrid beamforming to approach the performance of fully digital beamforming depends mainly on the number of RF chains, rather than the number of transmit antennas.
However, the average spectral efficiency of DAM based on hybrid beamforming with the fully connected structure is slightly lower than that based on fully digital beamforming. This is attributed to the unit modulus constraint imposed on the analog beamforming matrix in the hybrid scheme. Furthermore, the average spectral efficiency of hybrid beamforming with the partially connected structure is clearly inferior to that with the fully connected structure, which is due to the fact that the partially connected structure restricts each RF chain to be connected to a subset of antennas, reducing the degrees of freedom in beamforming. However, while hybrid beamforming suffers from some performance losses compared to fully digital beamforming, it offers a trade-off by reducing cost and hardware complexity.

%

\begin{figure}[!t]
	\centering
	\subfloat[Integer delays\label{BER_DAM-ID}]{
		\includegraphics[width=6.5cm]{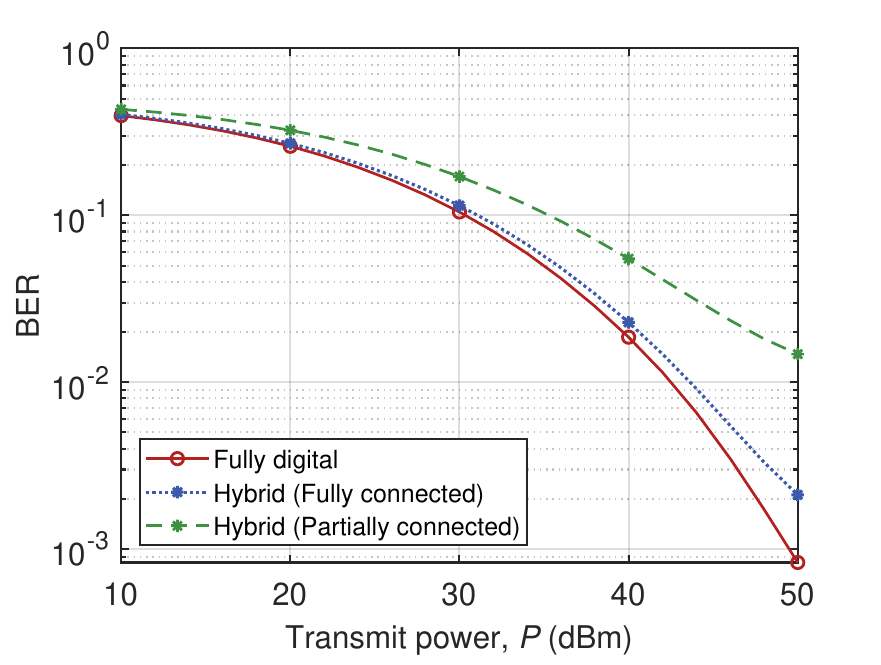}}\\
	\subfloat[Fractional delays\label{BER_DAM-FD}]{
		\includegraphics[width=6.5cm]{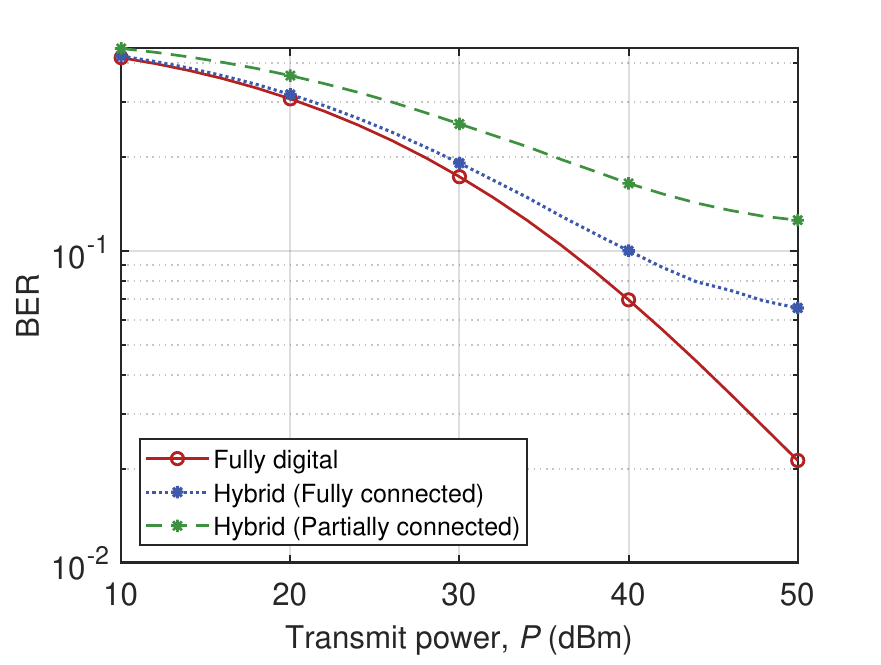}}
	\caption{BER performance of DAM for channels with integer delays and fractional delays.}
	\label{BER_DAM-IDFD}
\end{figure}

Fig.~\ref{BER_DAM-IDFD} shows the bit error rate (BER) performance comparison of DAM based on fully digital beamforming versus hybrid beamforming for channels with integer delays as well as fractional delays, respectively. The simulation results are obtained based on Monte Caro simulations. Specifically, we simulate the transmission process of random bit sequences of length $10^5$ using 128-QAM over the randomly generated channels $10^3$ times, and the BER is obtained by taking the average over all transmissions. The number of antennas is $M_\mathrm{t} = 256$. Fig.~\ref{BER_DAM-IDFD} reveals that, when the number of RF chains is $N_\mathrm{RF} = 4$, which is equal to that of multi-paths, the BER performance of DAM based on hybrid beamforming with fully connected structure is quite close to that based on fully digital beamforming, while the BER performance of that based on hybrid beamforming with partially-connected structure is clearly worse. Furthermore, comparing Fig.~\ref{BER_DAM-ID} and Fig.~\ref{BER_DAM-FD}, it is apparent that DAM performs worse in channels with fractional delays than in channels with integer delays. 
This is due to the strong correlation between adjacent taps, which leads to ISI that cannot be completely eliminated, resulting in the occurrence of a BER floor as the transmit power increases.

\begin{figure}[t!]
	\centerline{\includegraphics[width=6.5cm]{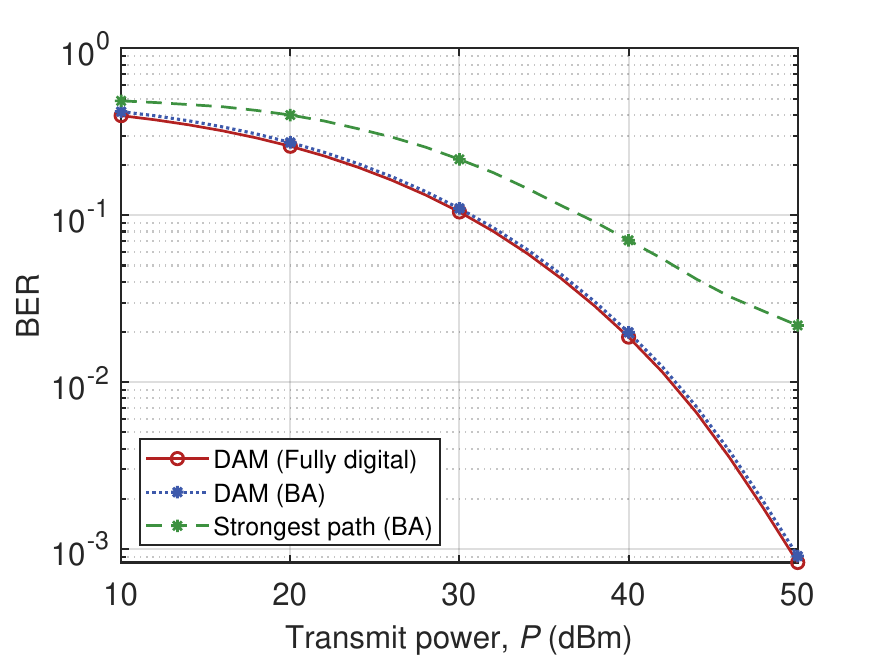}}
	\caption{BER comparison for DAM with beam alignment and fully digital beamforming, as well as the benchmarking strongest path scheme.}
	\label{BER_BA-DAM}
\end{figure}

Fig.~\ref{BER_BA-DAM} compares the BER performance of DAM with beam alignment (BA) and fully digital beamforming, as well as the benchmarking strongest path scheme. The number of RF chains used here is 16, and the fully connected structure is considered. It can be seen that the performance of DAM with beam alignment is identical to that of DAM with fully digital beamforming. This validates the efficacy of beam alignment design for DAM, indicating its practical application in mmWave/THz massive MIMO communication scenarios. It can also be clearly seen from Fig.~\ref{BER_BA-DAM} that the performance of benchmarking strongest path is worse than DAM, which is due to the fact that it only utilizes one channel tap, failing to fully utilize the channel or losing diversity, verifying that the DAM brings performance advantages due to its ability to fully utilize the channel power contributed by all multi-path components.

\begin{figure}[t!]
	\centerline{\includegraphics[width=6.5cm]{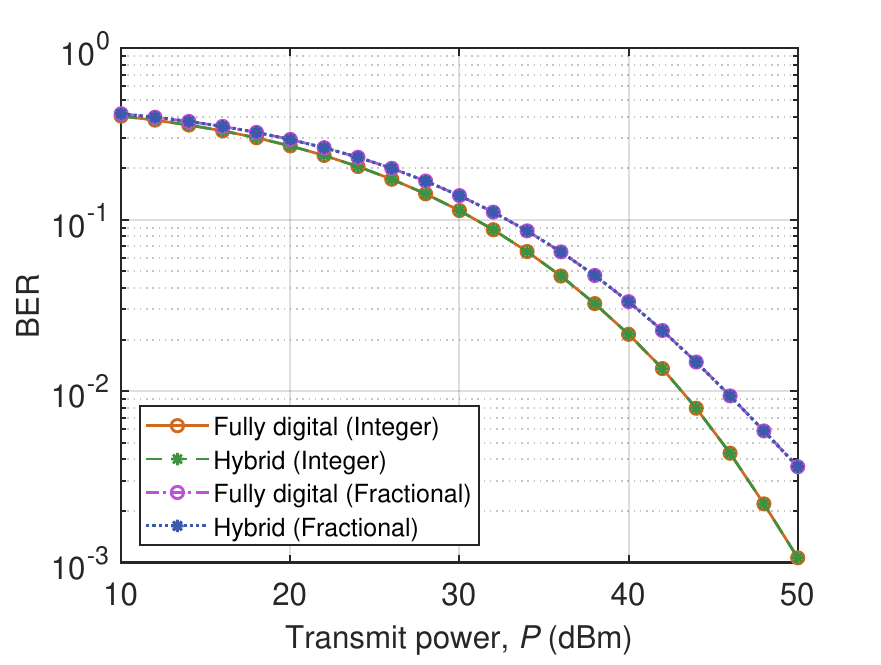}}
	\caption{BER comparison of DAM-OFDM based on fully digital beamforming versus hybrid beamforming for channels with integer delays as well as fractional delays.}
	\label{BER_DAM-OFDM}
\end{figure}

Fig.~\ref{BER_DAM-OFDM} presents the BER performance comparison of DAM-OFDM based on fully digital beamforming versus hybrid beamforming for channels with integer delays as well as fractional delays. The number of RF chains used here is 16, and it can be seen that when the number of RF chains is large enough, the performance of DAM-OFDM based on hybrid beamforming is almost equal to that based on fully digital beamforming. It can also be seen from Fig.~\ref{BER_DAM-OFDM} that, unlike what is presented in Fig.~\ref{BER_DAM-IDFD}, there is only a small degradation in the performance of DAM-OFDM over the channel with fractional delays as compared to that with integer delays. This is due to the fact that OFDM is able to further overcome ISI after DAM reduces the delay spread. The slight degradation is due to the correlation between different taps of the channel with fractional delays, which causes some performance loss in DAM with ZF beamforming. Nevertheless, the reduced delay spread due to DAM allows for a reduction in the use of OFDM sub-carriers or CP~\cite{DAM-OFDM}.

\section{Conclusion} \label{Conclusion}

This paper proposed the hybrid analog/digital beamforming-based DAM for mmWave/THz communications, including fully connected and partially connected structures. 
It leverages the high spatial resolution of large antenna arrays and the sparsity of multi-paths in mmWave/THz channels to mitigate ISI, with much fewer RF chains than the number of antennas.
The beam codebook based beam alignment DAM was further proposed, which can reduce the overhead and complexity of channel estimation. 
Furthermore, for channels with fractional delays, the proposed DAM-OFDM with hybrid beamforming method leverages first DAM to reduce delay spread and then OFDM to cope with the residual ISI, thus significantly reducing the number of OFDM sub-carriers and CP length. 
Simulation results verified the effectiveness of the proposed methods.

\end{document}